\documentstyle{l-aa}

\def\spose#1{\hbox to 0pt{#1\hss}}
\def\simlt{\mathrel{\spose{\lower 3pt\hbox{$\mathchar"218$}}
     \raise 2.0pt\hbox{$\mathchar"13C$}}}
\def\simgt{\mathrel{\spose{\lower 3pt\hbox{$\mathchar"218$}}
     \raise 2.0pt\hbox{$\mathchar"13E$}}}
\def\ie{{\rm i.e. }}

\begin{document}
\thesaurus{11.03.5; 11.09.3; 11.05.2; 11.01.1}
\title{On the trend of [Mg/Fe] among giant
elliptical galaxies}
\author{Francesca Matteucci$^{1,2}$, Raffaella Ponzone$^{1,2}$, 
Brad K. Gibson$^{3}$}
\offprints{F. Matteucci}
\institute{
Dipartimento di Astronomia, Universit\`a di Trieste, Via G.B. Tiepolo, 11,
34131 Trieste, Italy \and
SISSA/ISAS, Via Beirut 2-4, 34014 Trieste, Italy \and
Mt. Stromlo and Siding Spring Observatories, Australian National University,
Weston Creek, Australia ACT 2611}
\maketitle
\begin{abstract}
We revisit the problem of the flat slope of the $Mg_2$ versus $<Fe>$
relationship found for nuclei of elliptical galaxies (Faber et al. 1992;
Worthey et al. 1992; Carollo et al. 1993; Davies et al. 1993),
indicating that the Mg/Fe ratio should increase with galactic luminosity
and mass.
We transform the abundance of Fe, as predicted by classic wind models 
and alternative models for the chemical evolution of elliptical galaxies,
into the metallicity indices $Mg_2$ and $<Fe>$, by means of the more recent
index calibrations and show that none of the current models for the chemical 
evolution of elliptical galaxies is able to reproduce exactly 
the observed slope of the
$<Fe>$ versus $Mg_2$ relation, although the existing spread 
in the data makes this comparison quite difficult.
In other words, we can not clearly discriminate between models predicting
a decrease (classic wind model) or an increase of such a ratio with 
galactic mass.
The reason for this resides in the fact that the available observations
show a large spread due mostly to the errors in the derivation
of the $<Fe>$ index. In our opinion this fact prevents us from drawing 
any firm conclusion on the behaviour of Mg and Fe in these galaxies.
Moreover, as already shown by other authors, one should be careful in 
deriving trends in the real abundances just from the
metallicity indices, since these latter depend also on other physical 
parameters than the
metallicity.
This is an important point since abundance ratios have been proven to
represent strong constraints for galaxy formation mechanisms.
\keywords{galaxy:evolution--nucleosynthesis}
\end{abstract}
\bigskip\bigskip

\section{Introduction}
Elliptical galaxies do not show the presence of HII regions and it is not
possible to resolve single stars in them in order to measure photospheric
abundances. Therefore, most of the
information on these objects is obtained from
their integrated properties: abundances are derived either through
colors or integrated spectra and in both cases the derived information
is a complicated measure of metallicity and age (the well known 
age-metallicity degeneracy).
The most common metallicity
indicators are $Mg_{2}$ and $<Fe>$, as originally defined in 
Faber et al. (1977; 1985).
Population synthesis techniques are adopted to analyze the integrated 
properties of ellipticals and to derive
an estimate of their real abundances.
Unfortunately, they contain several uncertainties residing
either in incomplete knowledge of stellar evolution or in deficiencies
in stellar libraries, as discussed in Charlot et al. (1996).
In recent years more and more population synthesis models 
(Bruzual and Charlot, 1993; Buzzoni et al., 1992; Bressan et al. , 1994;
Gibson, 1997; Bressan et al., 1996; Gibson and Matteucci, 1997; Tantalo et 
al., 1998) 
have appeared
but the basic uncertainties still remain.
In this paper we want to focus our attention about the 
comparison between theoretical model results and metallicity indicators.
In this framework we will analyze the relationship between $<Fe>$ and $Mg_2$
and its implications for the mechanism of galaxy formation.

Several authors  (Faber et al. 1992; Worthey et al. 1992; 
Carollo et al. 
1993; 
Davies et al.
1993; Carollo and Danziger 1994),
from comparison of the observed indices with synthetic indices, concluded that
the average $[<Mg/Fe>]_{*}$ in giant ellipticals must be larger 
than the solar value.
This result was also confirmed by the analysis of Weiss et al. (1995) 
who made use, for the first time, 
of stellar evolutionary tracks calculated under the assumption 
of non-solar ratios.The same authors found that the $<Fe>$ versus $Mg_{2}$
relation among nuclei of giant ellipticals is rather flat and flatter 
than within galaxies. 
From the flat behavior of $<Fe>$ vs. $Mg_2$ the same authors
inferred that the abundance of Mg should increase 
faster than the abundance of Fe among nuclei of giant ellipticals.
This conclusion is at variance with the predictions of supernova-driven 
wind models of ellipticals (Arimoto and Yoshii, 1987;
Matteucci and Tornamb\`e 1987).
In fact, Matteucci and Tornamb\`e (1987) showed that,
in the framework of the classic wind model for ellipticals, 
the [Mg/Fe] ratio is a decreasing function of the galactic mass
and luminosity. The reason for this behavior is  clear: if Fe is mostly 
produced by the supernovae of type Ia, as it seems to be  the case 
in our Galaxy 
(Greggio and Renzini 1983a; Matteucci and Greggio 1986), 
whereas Mg is mostly originating from 
supernovae of type II,
then the iron production is delayed relative to that of Mg and
its abundance should be larger
in more massive galaxies which develop a wind 
later than the less massive ones.
All of this is valid
under the assumption that after the onset of a galactic wind 
star formation should stop or should be negligible, which is a reasonable 
assumption for elliptical galaxies.
Faber et al. (1992) proposed alternative scenarios to 
the classic supernova driven wind model, as 
originally proposed by Larson (1974).
They suggested three different scenarios all based on the assumption that
Mg is produced by type II supernovae and Fe is mostly
produced by type Ia supernovae:
i) a selective loss of metals, ii) a variable initial mass function (IMF)
and 
iii) different timescales for star formation.
These hypotheses 
were discussed by Matteucci (1994), who tested them in the context of
chemical evolution models.
In the hypothesis of the  different timescales for star formation
Matteucci (1994)
suggested that the more massive ellipticals might
experience a much stronger and 
faster star formation than less massive ellipticals leading to a 
situation where the
most massive objects are able to develop galactic winds before the less 
massive ones. She called this case ``inverse wind model''.
On the other hand,
in the classic wind model of Larson (1974) the efficiency of star formation 
was the same for all
galaxies thus leading to the fact that the galactic wind in more massive 
systems occurs later than in less massive ones,
due to their deeper potential well.
In the models of Arimoto and Yoshii (1987) and Matteucci and Tornamb\`e (1987)
the efficiency of star formation was a decreasing function of galactic mass, 
based on the assumption that the timescale for star formation is proportional
to the cloud-cloud collision timescale which, in turn, is proportional to
the gas density. Therefore, since in this monolithic collapse picture
the gas density decreases with the galactic mass,
the galactic wind 
was even more delayed for the most massive systems.
Matteucci (1994) proposed, as an alternative, a star formation efficiency 
increasing with the 
galactic mass and she justified this assumption by imagining 
that giant elliptical galaxies, instead 
of forming through a monolithic collapse of a gas cloud,  
form by
merging of gaseous protoclouds. 
The merging process can, in fact, produce
higher densities  for increasing galactic mass
and/or higher cloud-cloud collision velocities 
resulting
in a faster star formation process.
In such a model the galactic wind occurs earlier in massive than in smaller 
ellipticals thus producing the expected trend of an increasing [Mg/Fe] as a 
function
of galactic mass. 
Matteucci (1994) also showed that a variable IMF with the slope 
decreasing with increasing galactic mass and luminosity
can produce the same effect without an inverse wind situation.
The reason for that resides in the fact that a flatter IMF slope favors massive stars relative to low and intermediate masses, thus favoring Mg production
over Fe production.
However, Matteucci (1994) could not translate the predicted abundances 
into $Mg_2$ and $<Fe>$ since there were no 
available calibrations for [Fe/H] versus $<Fe>$ but only calibrations 
for $[Fe/H]$ vs. $Mg_2$. Therefore she did not compare the 
predicted abundances with observations.

Recently, calibrations for the iron index have become 
available (Borges et al., 1995; 
Tantalo et al., 1998) and therefore
in this paper we revisit the whole problem of inferring trends on the real
abundances by metallicity indices and we discuss the influence of the 
calibration relationships, which allow us to pass from indices 
to abundances, and 
we show that the inferred trend of Mg/Fe with galactic mass 
is not so clear when interpreted in terms of real abundances, 
thus warning us from drawing 
any firm conclusion on galaxy formation processes just on the basis of 
the observed behavior of $<Fe>$ versus $Mg_{2}$.
indices. The reason for that resides partly in 
the large spread present in the observational data and partly in
the fact 
that metallicity indices 
depend not only on the abundances of single elements but also on the ages 
and on the metallicity distribution (Tantalo et al. 1998) of the
different stellar populations present in elliptical galaxies.
\par
In Section 2
we will discuss the chemical evolution model, in Section 3 
we will define the average abundances of a composite stellar population,
in Section 4 we will describe the model results and transform the
predicted abundances into indices by means of the most recent
metallicity calibrations.
Finally in Section 5 some conclusions will be drawn.

\noindent
\section{The chemical evolution model}
\subsection{Basic equations}
\noindent
The adopted model of galactic evolution is that outlined by
Matteucci and Gibson (1995)(hereafter MG95),
where extensive descriptions and references can be found. The
evolution of the abundances of several chemical species (H, He, C, N, O, Ne,
Mg, Si and Fe) in the gas is followed, taking into account
detailed nucleosynthesis from stars of all masses and SNe of types Ia, Ib, and
II. 
We
assume that ellipticals can be considered initially as homogeneous spheres of
gas with luminous mass in the range $10^{9} \rightarrow 1\times 10^{12}$
M$_\odot$.  A single zone interstellar medium (ISM) with instantaneous mixing
of gas is assumed throughout.  
The adopted age for all
galaxy models is $t_{\rm G} = 15$ Gyr.

\par
The fundamental equations can be written
\small
\begin{eqnarray}
\lefteqn{{dG_i(t) \over dt} = -\psi(t)X_i(t)+\int_{M_{\rm L}}^{M_{\rm Bm}}
{\psi(t-\tau_m)Q_{mi}(t-\tau_m)\phi(m){\rm d}m}} \nonumber \\
& & +A\int_{M_{\rm Bm}}^{M_{\rm BM}}{\phi(m)}\biggl[\int_{\mu_{\rm min}}
^{0.5}{f(\mu)\psi(t-\tau_{m2})Q_{mi}(t-\tau_{m2}){\rm d}\mu\biggr]{\rm d}m}
\nonumber \\
& & +(1-A)\int_{M_{\rm Bm}}^{M_{\rm BM}}{\psi(t-\tau_{m})Q_{mi}(t-\tau_m)\phi(m)
{\rm d}m} \nonumber \\
& & +\int_{M_{\rm BM}}^{M_{\rm U}}{\psi(t-\tau_m)Q_{mi}(t-
\tau_m)\phi(m){\rm d}m}, \label{eq:evolution}
\end{eqnarray}
\normalsize
\noindent
where $G_i(t)=\rho_{\rm gas}(t)X_i(t)/\rho(0)$ is the volume gas density in the
form of an element {\it i} normalized to the initial total volume gas density.
The quantity $X_i(t)$ represents the abundance by mass of an element {\it i}
and by definition the summation over all the elements present in the gas
mixture is equal to unity.\par
\par
The various integrals in equation (1)
represent the rates at which SNe (I and II) as
well as single low and intermediate mass stars and single massive stars restore
their processed and unprocessed material into the ISM (for a detailed 
description of the integrals see MG95;
Matteucci and Greggio 1986). 
We only remind here that
the quantity $Q_{mi}$
represents the fraction of a star of mass $m$ which is restored in to the ISM in
the form of an element $i$
and therefore contains the nucleosynthesis
prescriptions that we assume to be the same as in MG95.

\par
The star formation rate $\psi(t)$ is given by:

\begin{equation}
\psi(t) = \nu\rho_{\rm gas}(t)/\rho(0).  \label{eq:sfr}
\end{equation}
\noindent

\ie normalized to the initial total volume density.  $\psi(t)$ is
assumed to drop
to zero at the onset of the galactic wind. The quantity $\nu $ is expressed in
units of Gyr$^{-1}$ and represents the efficiency of star formation, namely the
inverse of the time scale of star formation.  The values adopted here for $\nu$
are the same as in Matteucci (1992) and Arimoto and Yoshii (1987). 
In one case we
have adopted the prescription for the {\bf inverse wind model} of Matteucci
(1994). The difference between the two cases is that in Matteucci (1994)
the {\bf classic wind model} assumes that the efficiency of star formation
decreases with increasing total galactic mass,
as in Arimoto and Yoshii (1987),
whereas in the inverse wind model
the efficiency of star formation
increases as the total galactic mass increases (similar in spirit to the
SFR efficiency parametrization of Tinsley and Larson 1979)
leading to a situation in which
more massive galaxies experience a galactic wind before the less massive ones.
\par
The quantity $\tau_m$ represents
the lifetime of a star of mass $m$, and is taken
from Padovani and Matteucci (1993).
\subsection{Galactic winds}
\noindent

For gas to be expelled from a
galaxy the following condition should be
satisfied: the thermal energy of the gas heated by SN explosions 
should exceed
the binding energy of the gas (Larson 1974).  At this point the 
gas present in
the galaxy is swept away and the subsequent evolution is determined 
only by the
amount of matter and energy which is restored to the ISM by the dying stellar
generations. In particular, only low mass stars contribute to 
this evolutionary
phase and, among the SNe, only SNe of type Ia (\ie those SNe events whose
progenitors have the longest lifetimes).
\par
Therefore, in order to evaluate the time for the onset of a galactic wind we
need to know the energy input from SNe and the binding energy of the gas as a
function of time. The total thermal energy of the gas at time t,
$E_{\rm th_{\rm SN}}$, and
the binding
energy of the gas in presence of a diffuse halo of dark matter are calculated
as described in Matteucci (1992).
In particular, $E_{\rm th_{\rm SN}}(t)$ is calculated by assuming
that $\sim 70\%$ of the initial blast wave energy is transferred into the ISM
as thermal energy by a SN remnant, if the time elapsed from the SN explosion
is shorter than a SN remnant cooling time (Cox 1972). The percentage of
transferred energy then decreases as a power law in time
$\propto t^{-0.62}$ for times larger than the cooling time.

This is the same formulation used by Arimoto and Yoshii (1987),
Matteucci and Tornamb\`e (1987)
and MG95 with the exception that we consider 
a more realistic cooling time (expressed in years)
which takes into account that the gas density
is changing with time and that part of the interstellar gas is
in the form of He:
\begin{eqnarray}
t_c = 5.7 \cdot 10^{4}[2.99 \cdot 10^{-3}(M_{gas}(t)/10^{12}M_{\odot})
\cdot \nonumber \\ 
(M_{lum}/10^{12}M_{\odot})^{-1.65}]^{-0.53}\,\, years. \label{eq:tc}
\end{eqnarray}

Other formulations of
the energy input from SNe can be found in
Gibson (1997) and Gibson and Matteucci (1997). 
\par
The energetic input from stellar winds in massive stars is ignored,
since for normal ellipticals it is
negligible compared to the SN thermal energy contribution,
as shown by Gibson (1994). 

\par
The galactic wind may last only for few $10^{8}$ years or continue until 
the present time depending crucially on the assumptions about 
the thermal energy of the gas
and the potential energy of the gas. Unfortunately, none of these quantities 
is well known.
The time for the occurrence of the galactic wind, $t_{\rm GW}$, 
either increases with
the galactic mass as a consequence of the potential 
well incresing with galactic mass together with
the efficiency
of star formation decreasing with galactic mass (classic wind model), 
or decreases with the galactic mass
if the efficiency of star formation is strongly increasing 
with galactic mass (inverse wind model),
as can be seen in Table 3, as will be discussed in Section 4.\par
\par
\begin{table*}
\caption[]{Model I-classic wind, x=1.35}
\begin{flushleft}
\begin{tabular}{lllll}
\noalign{\smallskip}
\hline
\noalign{\smallskip}
$M_{lum}$($M_\odot$) & $\nu$\,($Gyr^{-1}$) & $R_{eff}$\,(kpc) &
$t_{GW}$\,(Gyr) & $M_{fin}$\,($M_\odot$)\\
\noalign{\smallskip}
\hline\noalign{\smallskip}
$10^9$   & 19.0 &  0.5   & 0.131  & $0.658\cdot10^9$\\ 
$10^{10}$  & 14.6 &  1.0   & 0.286  & $0.791\cdot10^{10}$\\ 
$10^{11}$  & 11.2 &  3.0   & 0.514  & $0.974\cdot10^{11}$\\ 
$10^{12}$  & 8.6  &  10.0  & 0.955  & $0.885\cdot10^{12}$\\ 
\noalign{\smallskip}
\hline
\end{tabular}
\end{flushleft}
\end{table*}
\begin{table*}
\caption[]{Model II- classic wind, x=0.95}
\begin{flushleft}
\begin{tabular}{lllll}
\noalign{\smallskip}
\hline
\noalign{\smallskip}
$M_{lum}$($M_\odot$) & $\nu$\,($Gyr^{-1}$) & $R_{eff}$\,(kpc) &
$t_{GW}$\,(Gyr) & $M_{fin}$\,($M_\odot$)\\ 
\noalign{\smallskip}
\hline\noalign{\smallskip}
$10^9$   & 19.0 &  0.5   & 0.069  & $0.246\cdot10^9$  \\
$10^{10}$  & 14.6 &  1.0   & 0.117  & $0.388\cdot10^{10}$\\
$10^{11}$  & 11.2 &  3.0   & 0.403  & $0.668\cdot10^{11}$\\ 
$10^{12}$  & 8.6  &  10.0  & 0.660  & $0.913\cdot10^{12}$\\ 
\noalign{\smallskip}
\hline
\end{tabular}
\end{flushleft}
\end{table*}
\begin{table*}
\caption[]{Model III- inverse wind, x=0.95}
\begin{flushleft}
\begin{tabular}{lllll}
\noalign{\smallskip}
\hline
\noalign{\smallskip}
$M_{lum}$($M_\odot$) & $\nu$\,($Gyr^{-1}$) & $R_{eff}$\,(kpc) &
$t_{GW}$\,(Gyr) & $M_{fin}$\,($M_\odot$)\\ 
\noalign{\smallskip}
\hline\noalign{\smallskip}
$10^9$   & 2. &  0.5   & 1.200  & $0.502\cdot10^9$  \\
$10^{10}$  & 5. &  1.0   & 0.897  & $0.605\cdot10^{10}$\\
$10^{11}$  & 11. &  3.0   & 0.408  & $0.665\cdot10^{11}$\\ 
$10^{12}$  & 20.  &  10.0  & 0.205  & $0.841\cdot10^{12}$\\
\noalign{\smallskip}
\hline
\end{tabular}
\end{flushleft}
\end{table*}
\begin{table*}
\caption[]{Model IV-classic wind, variable IMF}
\begin{flushleft}
\begin{tabular}{llllll}
\noalign{\smallskip}
\hline
\noalign{\smallskip}
$M_{lum}$($M_\odot$) & $\nu$\,($Gyr^{-1}$) & $R_{eff}$\,(kpc) &
$t_{GW}$\,(Gyr) & $M_{fin}$\,($M_\odot$) & IMF slope\\
\noalign{\smallskip}
\hline\noalign{\smallskip}
$10^9$   & 19.0 &  0.5   & 0.163  & $0.936\cdot10^9$  & 2.0  \\ 
$10^{10}$  & 14.6 &  1.0   & 0.237  & $0.945\cdot10^{10}$ & 1.7\\
$10^{11}$  & 11.2 &  3.0   & 0.514  & $0.974\cdot10^{11}$ & 1.35\\
$10^{12}$  & 8.6  &  10.0  & 0.660  & $0.912\cdot10^{12}$ & 0.95\\ 
\noalign{\smallskip}
\hline
\end{tabular}
\end{flushleft}
\end{table*}

\begin{table*}
\caption[]{Model V- classic wind, time variable IMF}
\begin{flushleft}
\begin{tabular}{lllll}
\noalign{\smallskip}
\hline
\noalign{\smallskip}
$M_{lum}$($M_\odot$) & $\nu$\,($Gyr^{-1}$) & $R_{eff}$\,(kpc) &
$t_{GW}$\,(Gyr) & $M_{fin}$\,($M_\odot$) \\
\noalign{\smallskip}
\hline\noalign{\smallskip}
$10^9$   & 19.0 &  0.5   & 0.073  & $0.276\cdot10^9$  \\ 
$10^{10}$  & 14.6 &  1.0   & 0.223  & $0.764\cdot10^{10}$\\
$10^{11}$  & 11.2 &  3.0   & 0.601  & $0.960\cdot10^{11}$\\
$10^{12}$  & 8.6  &  10.0  & 2.205  & $0.995\cdot10^{12}$\\ 
\noalign{\smallskip}
\hline
\end{tabular}
\end{flushleft}
\end{table*}

\begin{table*}
\caption[]{Model VI- inverse wind, x=0.8}
\begin{flushleft}
\begin{tabular}{lllll}
\noalign{\smallskip}
\hline
\noalign{\smallskip}
$M_{lum}$($M_\odot$) & $\nu$\,($Gyr^{-1}$) & $R_{eff}$\,(kpc) &
$t_{GW}$\,(Gyr) & $M_{fin}$\,($M_\odot$) \\
\noalign{\smallskip}
\hline\noalign{\smallskip}
$10^9$   &  4.0 &  0.5   & 0.22   & $0.320\cdot10^9$  \\ 
$10^{10}$  & 18.0 &  1.0   & 0.110  & $0.430\cdot10^{10}$\\
$10^{11}$  & 32.0 &  3.0   & 0.090  & $0.550\cdot10^{11}$\\
$10^{12}$  & 75.0 &  10.0  & 0.030  & $0.450\cdot10^{12}$\\ 
\noalign{\smallskip}
\hline
\end{tabular}
\end{flushleft}
\end{table*}

\section{The average metallicity of a composite stellar population}

In order to compare model results with observations we should first calculate
the average abundances of Mg and Fe of the composite stellar population
of the given galaxies.
The average metallicity or abundance in general which should be compared
with the indices should be averaged on the visual light, namely:

\begin{equation}
<X_i>_{L}= {\sum_{ij}{n_{ij} X_i L_{V_{j}}} \over \sum_{ij}{n_{ij}L_{V_j}}}
\end{equation}

where $n_{ij}$ is the number of stars in the abundance interval $X_i$  and luminosity $L_{V_j}$.
On the other hand, the real average abundance should be the mass-averaged one, namely:
\begin{equation}
<X_i>_{M} = {1 \over S_1} \int^{S_1}_{0}{X_i(S)dS}
\end{equation}
where the subscript $1$ refers to the specific time $t_1$ 
(the present time) and $S_1$ is the total mass of stars ever born.
Here we will use eq. (5) in order to compare model results 
with indices and the reason is that we want
to compare our results with previous ones (Matteucci 1994;
MG95) and because for giant ellipticals the difference between the 
mass-averaged metallicity and the luminosity-averaged one is
negligible (Yoshii and Arimoto, 1987; Gibson 1997).
On the other hand, in smaller systems the difference between  the two abundances is not negligible due to
the contribution to the light of low metallicity red giants (Greggio 1996).
We will show in Section 4 that in the cases studied here $<X_i>_{L} \simeq
<X_i>_{M}$.

\section{Model results}

In this section we will show the results of several models and we will convert,
by means of the available calibrations,
the  average  stellar  abundances of Mg  and  Fe, 
$[<Mg/H>]_{*}$ and $[<Fe/H>]_{*}$, 
as predicted for elliptical galaxies 
of different mass, into 
the metallicity indices $Mg_{2}$ and $<Fe>$, respectively. 
\par
Let us discuss first the metallicity calibrations.
Relations linking the strength of metallicity indices to real abundances can
be either empirical or theoretical. In the past few years several attempts 
have been made to calibrate the strength of $Mg_{2}$ against [Fe/H] which
has always been considered as the measure of the ``metallicity'' in stars.
It should be said that this is not entirely correct since we know that 
Mg does not evolve in lockstep with iron in the solar neighborhood
nor in elliptical galaxies, 
due to the 
different timescales of production of these two elements.
It would be much better to calibrate $Mg_2$ versus [Mg/H]
in order to avoid confusion.
Calibrations of $Mg_2$ versus $<Fe>$
are from Mould (1978), Burstein (1979),
Peletier (1989),
Buzzoni et al. (1992), Worthey et al. (1992). 
In all of these calibrations the ratio [Mg/Fe]
is assumed to be solar, at variance with the indication arising
from population synthesis models showing an overabundance of 
Mg relative to iron in the
nuclei of giant ellipticals (Faber et al. 1992; Worthey et al. 1992; 
Davies et al. 1993; Weiss et al. 1995). 

More recently, Barbuy (1994),
Borges et al. (1995) and Tantalo et al. (1998) took into account
non solar ratios of [Mg/Fe] in their calibrations.
In addition, some of them (Borges et al. 1995; Tantalo et al. 1998)
produced synthetic $<Fe>$ indices thus allowing us to calibrate [Fe/H] 
also against $<Fe>$.
This allows us to transform [Fe/H] in to $Mg_2$ and $<Fe>$, although
many uncertainties are involved in this exercise, mainly because, in this way,
the derivations of  $Mg_2$ and $<Fe>$ are not independent.  

\par
We run several models, in particular:
Model I, which is the classic wind model, as described in MG95, with a 
Salpeter (1955) IMF (namely an IMF with power index x=1.35
over a mass range $0.1 \le M/M_{\odot} \le 100$); 
Model II, which is the classic wind model 
with the Arimoto and Yoshii (1987) IMF (namely an IMF with power index
x=0.95 over the same mass range of the Salpeter one); 
Model III, which is the equivalent 
of the inverse wind model, as described in Matteucci (1994) 
with the Arimoto and Yoshii (1987) IMF; Model IV which is the equivalent of 
the model with variable IMF, as described in Matteucci (1994), which assumes
that ellipticals of smaller mass have a steeper IMF than the more massive ones.
It is worth noting that the slope of the IMF is kept constant inside a
galaxy.
In particular, we vary the slope of the IMF from the Salpeter one to 
the Arimoto and Yoshii one passing from a galaxy with initial luminous 
mass of $10^{11}M_{\odot}$ to a galaxy with $10^{12}M_{\odot}$.
This particular assumption can reproduce the observed tilt of the 
fundamental plane seen edge-on, namely the increase of M/L versus L as 
observed by
Bender et al. (1992).\par
Model V assumes a time-variable IMF as suggested by Padoan et al. (1997)
and will be discussed in a forthcoming paper. In this formulation 
the IMF
slope varies as a function of gas density and gas velocity dispersion,
favoring the formation of massive stars at early epochs.\par
Model VI assumes a constant IMF with a slope x=0.8 and a star formation efficiency which varies more strongly with the luminous mass than in Model III.
The slope x=0.8 is the limiting slope that we can accept to obtain a realistic $M/L_B$ ratio for ellipticals, as discussed in Padovani and Matteucci (1993).
\par
The model parameters are described in Tables 1-6
where we list the luminous masses
in column 1, the star formation efficiency (in units of $Gyr^{-1}$) 
in column 2,
the effective radius (in units of Kpc) in column 3, the time for the
occurrence of the galactic wind (in Gyr) in column 4 and the final 
galactic luminous mass in column 5.
For model IV is shown also the slope of the IMF in column 6.
\par
\begin{table*}
\caption[]{Model I- classic wind, x=1.35}
\begin{flushleft}
\begin{tabular}{lllll}
\noalign{\smallskip}
\hline
\noalign{\smallskip}
$log\,M_{lum}$($M_\odot$) & $[<Fe/H>]_\ast$\,(dex) & $[<Mg/Fe>]_\ast$\,(dex) &
$Mg_2$\,(mag) & $<Fe>$\,({\rm \,\AA})\\ 
\noalign{\smallskip}
\hline\noalign{\smallskip}
$9.$  & -0.395 & 0.262  & 0.217  & $2.545$  \\ 
$10.$ & -0.190 & 0.224  & 0.247  & $2.853$\\ 
$11.$ & -0.095 & 0.187  & 0.257  & $2.990$\\
$12.$ & -0.007 & 0.130  & 0.260  & $3.112$\\ 
\noalign{\smallskip}
\hline
\end{tabular}
\end{flushleft}
\end{table*}

\begin{table*}
\caption[]{Model  II- classic wind, x=0.95}
\begin{flushleft}
\begin{tabular}{lllll}
\noalign{\smallskip}
\hline
\noalign{\smallskip}
$log\,M_{lum}$($M_\odot$) & $[<Fe/H>]_\ast$\,(dex) & $[<Mg/Fe>]_\ast$\,(dex) &
$Mg_2$\,(mag) & $<Fe>$\,({\rm \,\AA})\\ 
\noalign{\smallskip}
\hline\noalign{\smallskip}
$9.$  & -0.833 & 0.397  & 0.152  & $1.937$  \\ 
$10.$ & -0.034 & 0.329  & 0.298  & $3.137$\\ 
$11.$ &  0.388 & 0.288  & 0.368  & $3.783$\\ 
$12.$ &  0.502 & 0.264  & 0.383  & $3.957$\\ 
\noalign{\smallskip}
\hline
\end{tabular}
\end{flushleft}
\end{table*}

\begin{table*}
\caption[]{Model  III- inverse wind, x=0.95}
\begin{flushleft}
\begin{tabular}{lllll}
\noalign{\smallskip}
\hline
\noalign{\smallskip}
$log\,M_{lum}$($M_\odot$) & $[<Fe/H>]_\ast$\,(dex) & $[<Mg/Fe>]_\ast$\,(dex) &
$Mg_2$\,(mag) & $<Fe>$\,({\rm \,\AA})\\ 
\noalign{\smallskip}
\hline\noalign{\smallskip}
$9.$  & 0.197 & 0.262  & 0.326  & $3.477$\\ 
$10.$ & 0.334 & 0.274  & 0.355  & $3.700$\\ 
$11.$ & 0.384 & 0.290  & 0.368  & $3.777$\\ 
$12.$ & 0.317 & 0.308  & 0.360  & $3.678$\\ 
\noalign{\smallskip}
\hline
\end{tabular}
\end{flushleft}
\end{table*}

\begin{table*}
\caption[]{Model  IV- classic wind, variable IMF}
\begin{flushleft}
\begin{tabular}{lllll}
\noalign{\smallskip}
\hline
\noalign{\smallskip}
$log\,M_{lum}$($M_\odot$) & $[<Fe/H>]_\ast$\,(dex) & $[<Mg/Fe>]_\ast$\,(dex) &
$Mg_2$\,(mag) & $<Fe>$\,({\rm \,\AA})\\ 
\noalign{\smallskip}
\hline\noalign{\smallskip}
$9.$  & -1.450 & 0.149  & 0.002  & $0.800$\\
$10.$ & -0.834 & 0.169  & 0.125  & $1.800$\\ 
$11.$ & -0.095 & 0.187  & 0.257  & $2.990$\\ 
$12.$ &  0.502 & 0.264  & 0.368  & $3.957$\\ 
\noalign{\smallskip}
\hline
\end{tabular}
\end{flushleft}
\end{table*}

\begin{table*}
\caption[]{Model  V- classic wind, time variable IMF}
\begin{flushleft}
\begin{tabular}{lllll}
\noalign{\smallskip}
\hline
\noalign{\smallskip}
$log\,M_{lum}$($M_\odot$) & $[<Fe/H>]_\ast$\,(dex) & $[<Mg/Fe>]_\ast$\,(dex) &
$Mg_2$\,(mag) & $<Fe>$\,({\rm \,\AA})\\ 
\noalign{\smallskip}
\hline\noalign{\smallskip}
$9.$  & -0.550 & 0.339  & 0.200  & $2.341$\\
$10.$ &  0.115 & 0.253  & 0.309  & $3.344$\\ 
$11.$ &  0.267 & 0.136  & 0.308  & $3.555$\\ 
$12.$ &  0.291 & -0.058  & 0.263  & $3.545$\\ 
\noalign{\smallskip}
\hline
\end{tabular}
\end{flushleft}
\end{table*}

\begin{table*}
\caption[]{Model VI- inverse wind, x=0.8}
\begin{flushleft}
\begin{tabular}{lllll}
\noalign{\smallskip}
\hline
\noalign{\smallskip}
$log\,M_{lum}$($M_\odot$) & $[<Fe/H>]_\ast$\,(dex) & $[<Mg/Fe>]_\ast$\,(dex) &
$Mg_2$\,(mag) & $<Fe>$\,({\rm \,\AA})\\ 
\noalign{\smallskip}
\hline\noalign{\smallskip}
$9.$  & -0.099 & 0.340  & 0.287  & $3.040$\\
$10.$ & -0.091 & 0.365  & 0.294  & $3.064$\\ 
$11.$ & -0.023 & 0.368  & 0.308  & $3.168$\\ 
$12.$ & -0.005 & 0.390  & 0.316  & $3.188$\\ 
\noalign{\smallskip}
\hline
\end{tabular}
\end{flushleft}
\end{table*}

We then calculate the average $[<Fe/H>]_{*}$ and $[<Mg/Fe>]_{*}$
for
the stellar component of ellipticals by using eq. (6) and  
finally we transform these abundances to observed 
$Mg_2$ and $<Fe>$ line indices. 
In Tables 7-12 we show the results for different models
and for the calibration of Tantalo et al. (1998).
In particular, in Tables 7-12 we show the luminous mass in column 1, the 
$[<Fe/H>]_{*}$ in the second column, the $[<Mg/Fe>]_{*}$ in column 3
and in column 4 and 5 the $Mg_2$ and the $<Fe>$ indices, respectively.
As already said, only the calibrations of Tantalo et al. (1998) and Borges 
et al. (1995)
allow us to transform [Fe/H] into $<Fe>$ and therefore allow us to compare
model results with the data showing the
behavior of $Mg_2$ vs. $<Fe>$ among nuclei of giant ellipticals.
In particular, starting from the synthetic indices of Tantalo et al. (1998)
calculated for a fixed [Mg/Fe] and a fixed [Fe/H] we derived calibration 
relationships of the type:
\begin{equation}
Mg_2=f([Fe/H], [Mg/Fe])
\end{equation}
\begin{equation}
<Fe>=g([Fe/H], [Mg/Fe])
\end{equation}
which allowed us to derive the indices for any [Fe/H] and [Mg/Fe].
The calibrations we have adopted  are:

\begin{eqnarray}
Mg_2=0.233+0.217[Mg/Fe]+ \nonumber \\
+(0.153+0.120[Mg/Fe])\cdot [Fe/H]
\end{eqnarray}

\begin{eqnarray}
<Fe>=3.078+0.341[Mg/Fe]+ \nonumber \\
(1.654 - 0.307[Mg/Fe])\cdot [Fe/H]
\end{eqnarray}

\begin{figure}
\centering
\vspace{12.0cm}
\caption[]{\label{fig:fig 1}a) Predicted and observed metallicity indices.
The dotted line and open squares
represent the $<Fe>$ versus $Mg_2$ predicted by
Model I for galaxies of different masses, as indicated in Table 7. The dashed
line and stars represent the real abundances of Fe and Mg as predicted by 
Model I for galaxies of different masses and arbitrarily translated 
in the $<Fe>$ vs.
$Mg_2$ diagram. The error bars referring to the different data samples
are also shown.
b) Predicted and observed $Mg_2$ versus mass relation. The predictions 
are from Model I.}

\end{figure}

\begin{figure}
\centering
\vspace{12.0cm}
\caption[]{\label{fig:fig 2}a) The same as Fig. 1a
but relative to the predictions of
Model II. 
b) The same as Fig. 1b but relative to the predictions of Model II.}

\end{figure}

In Fig. 1 we show the metallicity indices obtained by means of the 
already mentioned calibrations compared with the data (Gonzalez 1993;
Worthey et al. 1992; Carollo and Danziger 1994a,b).
As one can easily see the data present a large spread, mostly due
to the uncertainties in deriving the $<Fe>$ indices.
In particular, in Fig 1a we show the observed and predicted
behavior
of $<Fe>$ vs. $Mg_2$ when Model I is adopted. 
The bestfit to the data implies
the following relation, $<Fe>=3.94Mg_2+1.83$, and is indicated in the figure.
However, the spread in the data is large and this prevents us from drawing 
strong 
conclusions about a possible trend.
The dotted lines in Fig. 1a represent the predictions of Model I obtained
by means of the calibrations described before and adopting the same 
[Mg/Fe] ratio as predicted by the models,
as one can see in Table 7.
The agreement with the observed trend is not so good, showing
that the slope of the predicted relation is steeper than 
that shown by the data
and that the predicted $Mg_2$ values do not cover the entire range in $Mg_2$.
This is mostly due to the assumed IMF since Model II, which assumes 
a flatter IMF, predicts values for $Mg_2$ which cover the whole 
range (see Fig. 2a).

In Fig. 1b we show the predicted and observed mass-metallicity ($Mg_2$)
relationship. The data are from Carollo et al. (1993).
The best-fit to these data indicate $Mg_2 = 0.02logM_{tot}+0.08$,
where $M_{tot}$ is the total galactic mass (dark+luminous). 
The classic wind model recovers 
the slope of the $Mg_2$ -- mass relation, but with a zeropoint offset
of $\Delta Mg_2 \simeq 0.05$ with respect to the observed distribution.

On the other hand, the classic wind model with the
Arimoto and Yoshii (1987) IMF (Model II), as shown in Fig.2b,
predicts a slope much steeper
than the observed one, although it agrees better than Model I with the
$<Fe>$ vs. $Mg_2$ relation shown in Fig. 2a. 
It is worth noting that the  Arimoto and Yoshii (1987)  IMF
well reproduces the abundances in the
intergalactic medium (MG95; Gibson 1997; Gibson and Matteucci 1997).
It is worth noting that in figures 1a and 2a and in all the others we 
show also the relation between real abundances predicted by our models. 
The relation between 
$[<Fe/H>]_{*}$ and
$[<Mg/H>]_{*}$, arbitrarily translated in the plot
of $<Fe>$ versus $Mg_2$, is indicated by the dashed lines. This is done 
only with the purpose of comparing the slope of the 
relation between real abundances with that 
of the relation between indices and they are very similar, 
indicating that the adopted calibration  
does not modify the predicted relation between Mg and Fe abundances.
One of the main reasons for that is the adopted calibration which accounts for
the right $[<Mg/Fe>]_{*}$ ratio for each galaxy.
\begin{figure}
\centering
\vspace{12.0cm}
\caption[]{\label{fig:fig 3} a) The same as Fig. 1a but 
for the predictions of Model III. b) The same as 
Fig. 1b but for the predictions of Model III.}

\end{figure}
\par

In Fig.3a,b we show the predictions of the inverse wind model 
of Matteucci (1994) which predicts a stellar $[<Mg/Fe>]_{*}$ increasing with 
galactic mass. 
The slope of the $<Fe>$ versus $Mg_2$ relation is in better agreement 
than in the previous
models, and the slope of the $Mg_2$ vs. mass relation is also 
acceptable although the absolute values of the indices are too high.
\par
In Fig.4a,b we show the results of Model IV with a variable IMF from
galaxy to galaxy,
which also predicts increasing $[<Mg/Fe>]_{*}$ ratios with galactic mass.
The agreement with the 
$<Fe>$ vs. $Mg_2$ data is marginally acceptable, 
but the slope of the mass-metallicity relation is too steep
and the predicted absolute values of $Mg_2$ are too low.
The 
low absolute values of $Mg_2$ are due to the fact
that we used slopes steeper than the Salpeter (1955) one for the 
less massive galaxies and such slopes are not suitable for elliptical
galaxies (see MG95) since they predict too low metallicities.
However, other numerical experiments, where we used a variable 
IMF but with flatter slopes for each galactic mass
(from x=1.4 in low mass galaxies to x=0.8 in high mass galaxies),
have shown that there is a negligible difference in the predicted $<Fe>$ vs.
$Mg_{2}$ relation while the mass-metallicity relation gets worse.

In Fig.5a,b we show the predictions of Model V
calculated with the time-variable IMF as suggested by Padoan et al. 
(1997) and adapted to elliptical
galaxies. 
The slope of this IMF is decreasing with time thus favoring
massive stars at early epochs. A similar although more complex formulation
of the Padoan et al. (1997) IMF has been recently adopted by 
Chiosi et al. (1998).
The model behaves like 
a classic wind model, in the sense that the galactic wind occurs first 
in small galaxies and later in the more massive ones. 
Concerning the predicted indices, Fig.5a shows that the $Mg_2$ decreases 
for massive objects, due to the fact that the IMF in these galaxies 
is less biased towards massive stars than in smaller systems.
This is, in turn, due to the fact that the slope of the IMF is inversely 
proportional to the gas density which is lower in more massive objects.
This model predicts a sort of bimodal behavior for the $Mg_2$ vs. mass
relation and it does not fit the data better than the other models. 
\par
Finally, in Fig.6a,b the predictions of Model VI are shown.
At variance with all the previous models, Model VI seems to reproduce well the
observed slope of the $<Fe>$ vs. $Mg_{2}$ relation as well as the $Mg_2$-
mass relation. The main problem with this model is the
fact that the predicted ranges of $Mg_2$ and $<Fe>$ are too narrow 
compared to the observations, especially the range of $<Fe>$.
Another potential problem of this model is the 
predicted $M/L_B$ ratio which is $\sim 30$ for each galaxy mass.
This is a high value for ellipticals unless one believes in a Hubble constant
$H_o=100 Km s^{-1} Mpc^{-1}$, as discussed in Padovani and Matteucci (1993).

\begin{figure}
\centering
\vspace{12.0cm}
\caption[]{\label{fig:fig 4}a) The same as Fig. 1a but for the predictions 
of Model IV. b) The same as Fig. 1b but for the predictions of Model IV.}

\end{figure}

\par
In Figs. 7  and 8 we show the plot of the mass-metallicity 
($Mg_2$) relation
as predicted by Model I and Model II,
obtained under different assumptions about the calibrating formula.
As one can see, some of the calibrations give similar results such as those of
Worthey et al. (1992), Casuso et al. (1996) and Buzzoni et al. (1992).
These calibrations have in common the use of a solar ratio for
[Mg/Fe] ([Mg/Fe]=0). On the other hand, the values of the indices obtained 
by using the calibrations of Barbuy (1994)
and Tantalo et al. (1998) which adopt non-solar ratios, differ from 
the others and between themselves. 
It is worth noting that the use of different calibrations may lead even to
different slopes for the $Mg_2$-$log M$ relationship.

\begin{figure}
\centering
\vspace{12.0cm}
\caption[]{\label{fig:fig 5}a) The same as Fig. 1a but for the predictions of
Model V. b) The same as Fig. 1b but for the predictions of Model V.}

\end{figure}

\begin{figure}
\centering
\vspace{12.0cm}
\caption[]{\label{fig:fig 6}a) The same as Fig. 1a but for the predictions of
Model VI. b) The same as Fig. 1b but for the predictions of Model VI.}

\end{figure}

One criticism that could in principle be moved to the results discussed 
before is that we adopted mass-averaged metallicities and not luminosity-averaged metallicities, as it should be the case.
In Fig. 9 we show the 
indices obtained by the luminosity- and mass-averaged metallicities calculated
for the results of Model II, when the calibration of Tantalo et al. (1998)
is applied. The luminosity-averaged metallicities, computed with the 
photometric model of Gibson (1997), are systematically slightly lower 
than the others and the difference is larger for smaller galaxies, 
as expected. However, the slope is the same in the 
two cases, showing that the use
of mass-averaged metallicity for this kind of analysis is quite justified.

\begin{figure}
\centering
\vspace{8.0cm}
\caption[]{\label{fig:fig 7} $Mg_2$-mass relationships as predicted by Model I
by using different calibrations, as indicated in Fig. 8. The best-fit to the data is also shown.}

\end{figure}

\begin{figure}
\centering
\vspace{8.0cm}
\caption[]{\label{fig:fig 8}$Mg_2$-mass relationships as predicted by Model II
by using different calibrations, as indicated in the figure. The best-fit to the data is also
shown.}

\end{figure}

\begin{figure}
\centering
\vspace{8.0cm}
\caption[]{\label{fig:fig 9}Predicted and observed $<Fe>$ versus $Mg_2$.
The predicted indices, relative to Model II, are calculated by averaging on the mass (stars) and on the visual luminosities (squares).}

\end{figure}

\section{Discussion and Conclusions}
In this paper we have discussed the relation between metallicity indices,
such as $Mg_2$ and $<Fe>$, and total mass  
in nuclei of ellipticals and its implications in terms of models 
of formation and evolution of elliptical galaxies.
\par
In order to do that we have transformed the 
average abundance of Fe 
in the composite stellar population of the galaxy,
as predicted by different models of chemical evolution,  
into $Mg_2$ and $<Fe>$ indices by means of the available calibrations.\par

We have shown the results of classic wind models for ellipticals, 
such as those discussed by Arimoto and Yoshii (1987) 
and Matteucci and Tornamb\`e (1987), as well as the results of models
with variable IMF from galaxy to galaxy and with galactic winds occurring 
first in the more massive systems, implying that these systems are older
than the less massive ones.
We have found that it is not possible to establish clearly which kind of model
should be preferred, first of all because of the large spread present
in the data. 

Moreover, little difference is found in the predicted indices of models
which predict a $[<Mg/Fe>]_{*}$ either increasing or decreasing 
with galactic mass, although the data seem to suggest an increase of this 
ratio with galactic mass larger than predicted by any of the models.

On this basis, the classic wind model can not be considered worse
than the other models.
Actually, the classic wind model with a flat constant IMF seems to be the
only one which can reproduce the whole range of the observed indices.
However, if we isolate the data from Gonzalez (1993) and do not 
consider the others, then
in order to reproduce the flat slope of the $<Fe>$ versus $Mg_2$ relation,
as given by the best-fit of the data,
one should assume that Fe among the nuclei of ellipticals is almost constant 
whereas Mg increases from less massive to more massive nuclei.
This is not achieved by any of the models presented here since it 
would require quite ``ad hoc'' assumptions especially concerning
the type Ia SNe.

From the numerical experiments performed in this paper we can say that
a model which explains at the same time the mass-metallicity and the
iron-magnesium relation requires an inverse wind situation, 
with a strong increase of the star formation efficiency with galactic mass
(i.e. Model VI),
rather 
than a variable IMF from galaxy to galaxy, and an IMF with a slope 
$x=0.8$. However, a model of this type is not able to reproduce the observed
ranges of $<Fe>$ and $Mg_2$.  
We have also calculated models where amount and concentration of
dark matter increases, compatibly with
the formulation of the potential energy of the gas, 
with decreasing galactic luminous mass (see Persic et al. 1996), 
with the net result of
obtaining an ``inverse wind'' situation.
The results are very similar to those of Model III.
Therefore, to obtain a better agreement with observations one 
should invoke also in this case
an increase of the star formation efficiency with galactic mass.
This would certainly flatten the $<Fe>$ vs. $Mg_2$ relation but it would
further shrink the ranges of the predicted indices.
In fact, both an increasing star formation efficiency and a decreasing importance of dark
matter with increasing luminous galactic mass can be viable solutions to achieve the
situation of more massive ellipticals being older than less massive ones.

\par
In conclusion,
it is quite important to establish the value of [Mg/Fe] from the observational
point of view since
abundance ratios, 
such as [Mg/Fe], represent an important diagnostic to infer ages in 
galaxies, due to the different timescales for the Mg and Fe production.
Generally, a high [Mg/Fe] is interpreted as a young age and the upper limit 
for the age is given
by the time at which the chemical enrichment from type Ia SNe starts 
to become important.
This timescale depends not only on the assumed progenitors of type Ia SNe
but also 
on the star formation history of the galaxy considered 
(see Matteucci 1997) and for giant
elliptical galaxies this timescale is of the order of
$t_{SNeIa}\sim 3-4 \cdot 
10^{8}$ years
and in any case it can not be larger than 1 Gyr also for smaller systems.
This is at variance with what stated by Kodama and Arimoto (1997) who claim, 
on the basis of results concerning our Galaxy (Yoshii et al. 1996),
that this timescale is 
of the order of 1.5-2.5 Gyr.
This is indeed true for our Galaxy where the star formation history 
has been quite different than in ellipticals and it had been 
already pointed out 
in Greggio and Renzini (1983b) and in Matteucci and Greggio (1986).
This is a quite important point, both for the
galactic chemical enrichment and for the
predictions about SN rates at high redshift.

Therefore, an enhanced [Mg/Fe] indicates that the process of galaxy 
formation must have been very fast thus favoring
a monolithic collapse scenario rather than a merging scenario.
In this framework, a [Mg/Fe] ratio higher in more massive ellipticals than in 
less massive ones could be interpreted as due to their faster formation
and evolution (see Matteucci 1994; Bressan et al. 1996).
\par
An independent way of estimating the ages of ellipticals, where for ages 
we intend the time elapsed from the last star formation event, is to study 
the $H_{\beta}$ index. This index is, in fact, related to the age of 
the dominant stellar population, since it gives a measure of the 
turn-off color and metallicity. It can therefore be used to solve the
age-metallicity degeneracy. Bressan et al. (1996), by analyzing the $H_{\beta}$
and other physical parameters in the sample of ellipticals observed
by Gonzalez (1993), concluded
that massive galaxies should have stopped forming stars before less 
massive ones,
in agreement with the results of the inverse wind model discussed here.
Finally, we would like to point out that both models with a Salpeter IMF
and a variable IMF have a potential problem in reproducing high [$\alpha$/Fe]
ratios in the intracluster medium (ICM), as shown by their low average
$[<\alpha/Fe>]_*$ ratios (see Tables 7-12).
Therefore, in agreement with MG95 and Gibson and Matteucci (1997) we conclude
that a flat IMF is required to explain the high [$\alpha$/Fe] ratios,
as found by ASCA observations (Mushotzsky 1994).

\begin{acknowledgements} One of us (F.M.) wants to thank F. Carbognani
for his help with the data fitting procedures.
We also like to thank the referee G. Worthey for his careful reading 
and his suggestions.
\end{acknowledgements}

{}


\begin{thebibliography}{}
\bibitem[] {}
Arimoto, N., Yoshii, Y. 1987, A.A. {\bf 173}, 23

\bibitem[]{}
Barbuy, B. 1994 Ap.J. {\bf 430}, 218

\bibitem[]{}
Bender, R., Burstein, D. \& Faber, S.M.: 1992, Ap.J. {\bf 399},462 

\bibitem[]{}
Bender, R., Ziegler, B., Bruzual, G. 1996, Ap.J. {\bf 463}, L51

\bibitem[]{}
Bertin, G., Saglia, R.P., Stiavelli, M. 1991 Ap.J. {\bf 384}, 423

\bibitem[]{}
Borges, A.C., Idiart, T.P., de Freitas-Pacheco, J.A.,
Thevenin, F. 1995 A.J. {\bf 110}, 2408
\bibitem[]{}
Bressan, A.,Chiosi, C., Fagotto, F. 1994 Ap.J. Suppl. {\bf 94}, 63

\bibitem[]{}
Bressan, A., Chiosi, C. Tantalo, R. 1996 A.A. {\bf 311}, 425

\bibitem[]{}
Brocato, E., Matteucci, F., Mazzitelli, I. \& Tornamb\`e, A.: 1990,
     Ap.J.  {\bf 349}, 458 
\bibitem[]{}
Bruzual, G., Charlot, S. 1993 Ap. J. {\bf 405}, 538
\bibitem[]{}
Burstein, D, 1979 Ap. J. {\bf232}, 74
\bibitem[]{}
Buzzoni, A. Gariboldi, G., Mantegazza, L. 1992 A.J. {\bf 103}, 1814
\bibitem[]{}
Carollo, C.M., Danziger, I.J., Buson, L.: 1993, MNRAS 
{\bf 265}, 553
\bibitem[]{}
Carollo, C.M., Danziger, I.J. 1994a, MNRAS {\bf 270}, 523
\bibitem[]{}
Carollo, C.M., Danziger I.J. 1994b, MNRAS {\bf 270}, 743
\bibitem[]{}
Casuso, E., Vazdekis, A., Peletier, R., Beckman, J.E. 1996 
Ap.J. {\bf 458}, 533
\bibitem[]{}
Charlot, S., Worthey, G., Bressan, A. 1996 Ap. J. {\bf 457}, 625
\bibitem[]{}
Chiosi, C., Bressan, A.,Portinari, L., Tantalo, R. 1998, A.A. in press
\bibitem[]{}
Cox, D.P. 1972 Ap.J. {\bf 178}, 159
\bibitem[]{}
Davies, R.L., Sadler, E.M. \& Peletier, R.: 1993, MNRAS {\bf
                 262}, 650
\bibitem[]{}
Faber, S.M., Worthey, G. \& Gonzalez J.J. 1992, 
                         in {\it IAU Symp.
                          n.149}, eds. B. Barbuy and A. Renzini, p. 255   
\bibitem[]{}
Faber, S.M., Burstein, D., Dressler, A. 1977, A.J. {\bf 82}, 941
\bibitem[]{}

Faber,S.M., Freiel, E.D., Burstein, D., Gaskell, C.M. 1985, Ap.J. Suppl. 
{\bf 57}, 711
\bibitem[]{} 
Gibson, B.., K. 1994 MNRAS {\bf 271}, L35
\bibitem[]{}
Gibson, B.K. 1997, MNRAS {\bf 290}, 471
\bibitem[]{}
Gibson, B.K., Matteucci, F., 1997. MNRAS, {\bf 291}, L8
\bibitem[]{}
Gonzalez, J.J.  1993 PhD Thesis, University of California, Santa Cruz
\bibitem[]{}
Greggio, L., 1996, MNRAS {\bf 285}, 151 
\bibitem[]{}
Greggio, L., Renzini, A. 1983a, in {\it ``The First Stellar
                          Generations''},  Mem. Soc. Astron. It., 
                          Vol. 54, p.311
\bibitem[]{}
Greggio, L., Renzini, A. 1983b A. A. {\bf 118}, 217
\bibitem[]{}
Kodama, T., Arimoto, N. 1997, A.A. {\bf 320}, 41
\bibitem[]{}
Larson, R.B. 1974 MNRAS {\bf 169}, 229
\bibitem[]{}
Matteucci, F. 1992 Ap.J. {\bf 397}, 32 
\bibitem[]{}
Matteucci, F. 1994 A.A. {\bf 288}, 57
\bibitem[]{}
Matteucci, F. Gibson, B.K. 1995 A.A. {\bf 304}, 11
\bibitem[]{}
Matteucci, F., Greggio, L.: 1986, A.A. {\bf 154}, 279
\bibitem[]{}
Mould, J.R. 1978 Ap.J. {\bf 220}, 434
\bibitem[]{}
Mushotzsky, R. 1994 in ``Clusters of Galaxies'' ed. F. Durret et al.
Editions Frontieres p.167
\bibitem[]{}
Nomoto, K., Thielemann, F. K. \& Yokoi, K.: 1984, Ap.J. {\bf
                             286}, 644
\bibitem[]{}
Padoan, P, Nordlund, A.P., Jones, B.J.T. 1997, MNRAS {\bf 288}, 145
\bibitem[]{}
Padovani, P., Matteucci, F. 1993 Ap.J. {\bf 416}, 26
\bibitem[]{}
Peletier, R., 1989, PhD Thesis, University of Groeningen 
\bibitem[]{}
Persic, M., Salucci, P., Stel, F. 1996 MNRAS {\bf 281}, 27
\bibitem[]{}
Renzini, A., Voli, M.: 1981, A.A. {\bf 94}, 175
\bibitem[]{}
Salpeter, E. E.: 1955, Ap.J. {\bf 121}, 161
\bibitem[]{}
Tantalo, R., Chiosi, C., Bressan, A. 1998 A.A. in press
\bibitem[]{}
Thielemann, F.K., Nomoto, K., Hashimoto, M. 1993 {\it Origin and Evolution
                        of the Elements}, ed. N. Prantzos et al., 
                        Cambridge Univ. Press p. 297
\bibitem[]{}
Tinsley, B.M., Larson, R.B. 1979 MNRAS {\bf 186}, 503 
\bibitem[]{}
Weiss, A., Peletier, R., \& Matteucci,F.: 1995 A.A. {\bf 296}, 73
\bibitem[]{}
Worthey, G. 1994 Ap.J. Suppl. {\bf 95}, 107
\bibitem[]{}
Worthey, G., Faber, S.M. \& Gonzalez, J.J.: 1992, Ap.J. {\bf
           398}, 73
\bibitem[]{}
Woosley, S.E., Weaver, T.A. 1986 Ann. Rev. Astron. Astrophys. {\bf 24}, 205
\bibitem[]{} 
Woosley, S. E.: 1987, in 
           {\it ``Nucleosynthesis and Chemical Evolution''}, 16th Saas 
           Fee Course, eds. B. Hauck et al. (Geneva: Geneva Observatory), p.1
\bibitem[]{}
Yoshii, Y., Arimoto, N. 1987 A.A. {\bf 188}, 13
\bibitem[]{} Yoshii, Y., Tsujimoto, T., Nomoto, K. 1996, Ap.J. {\bf 462}, 266

\end{thebibliography}
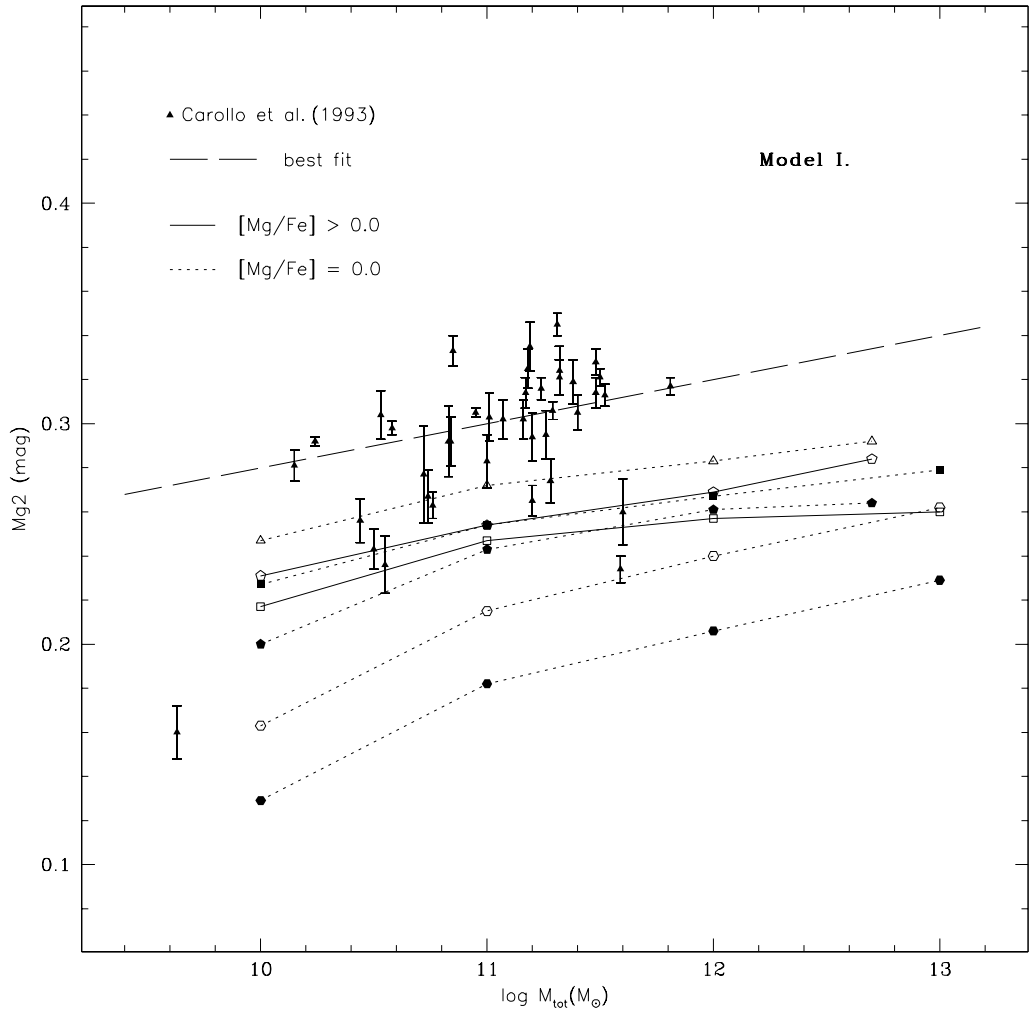
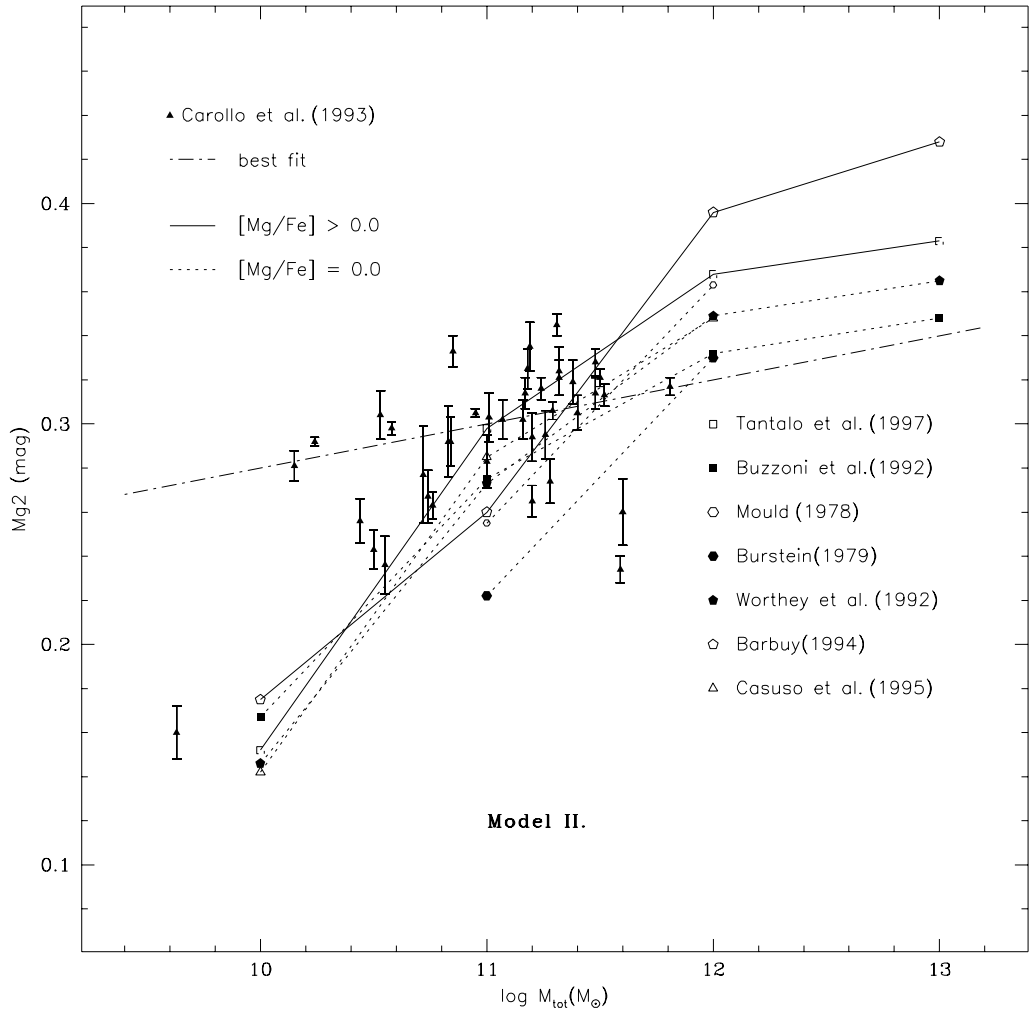
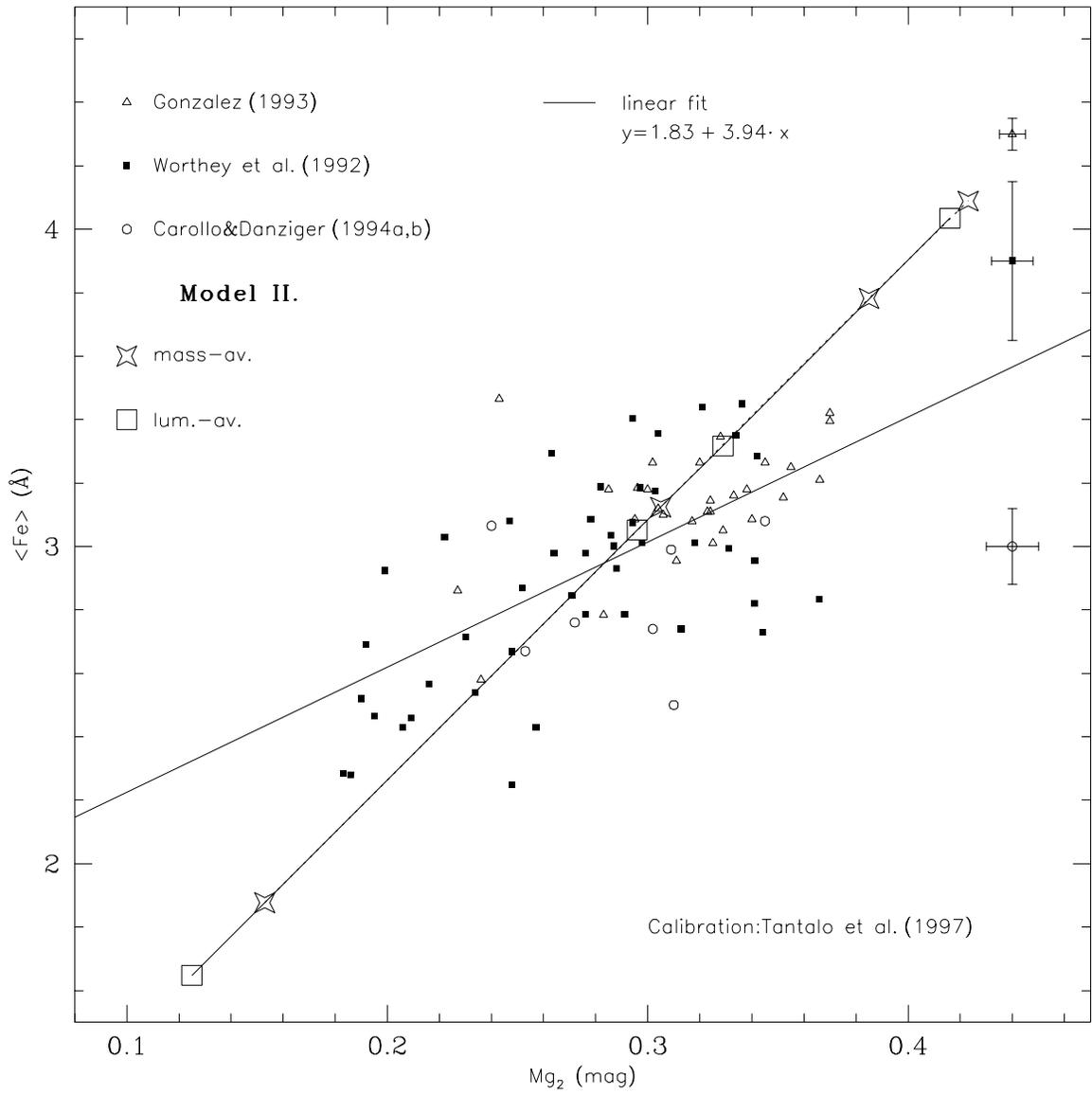
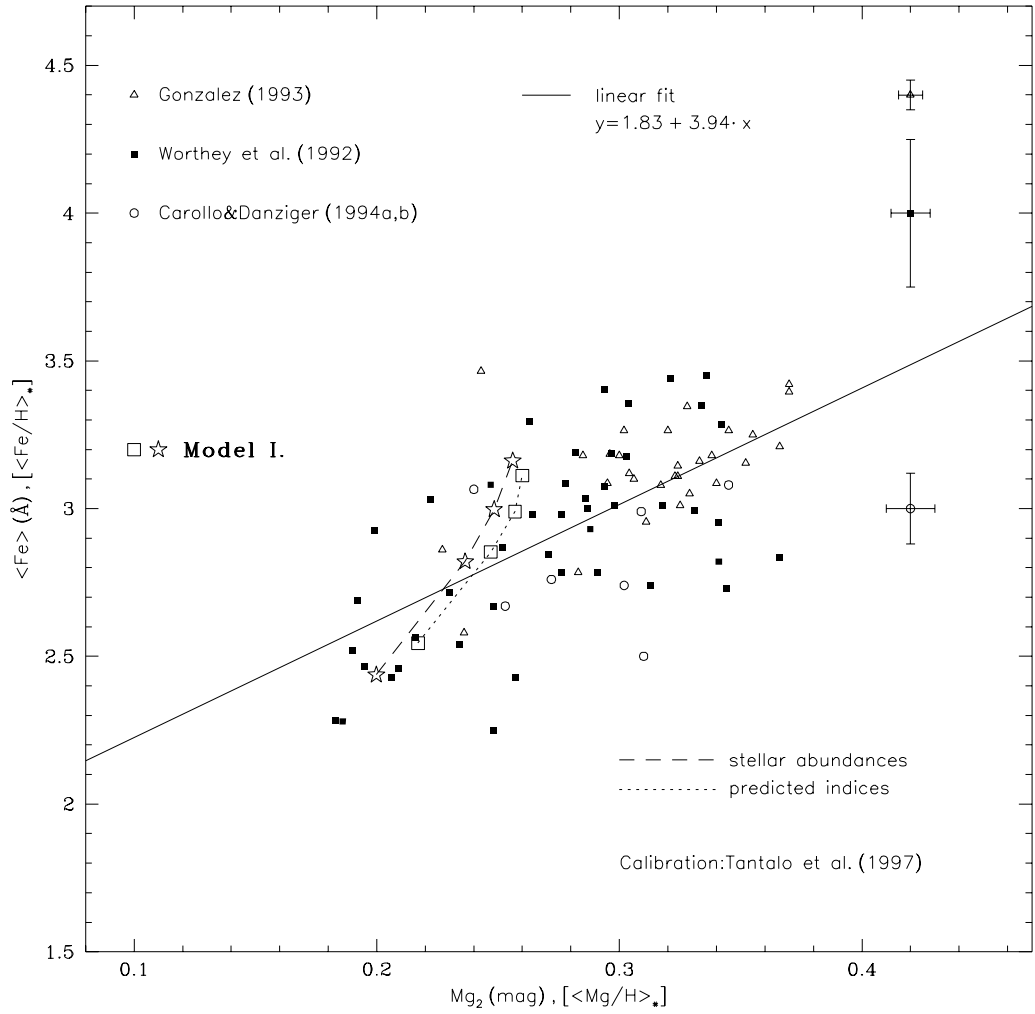
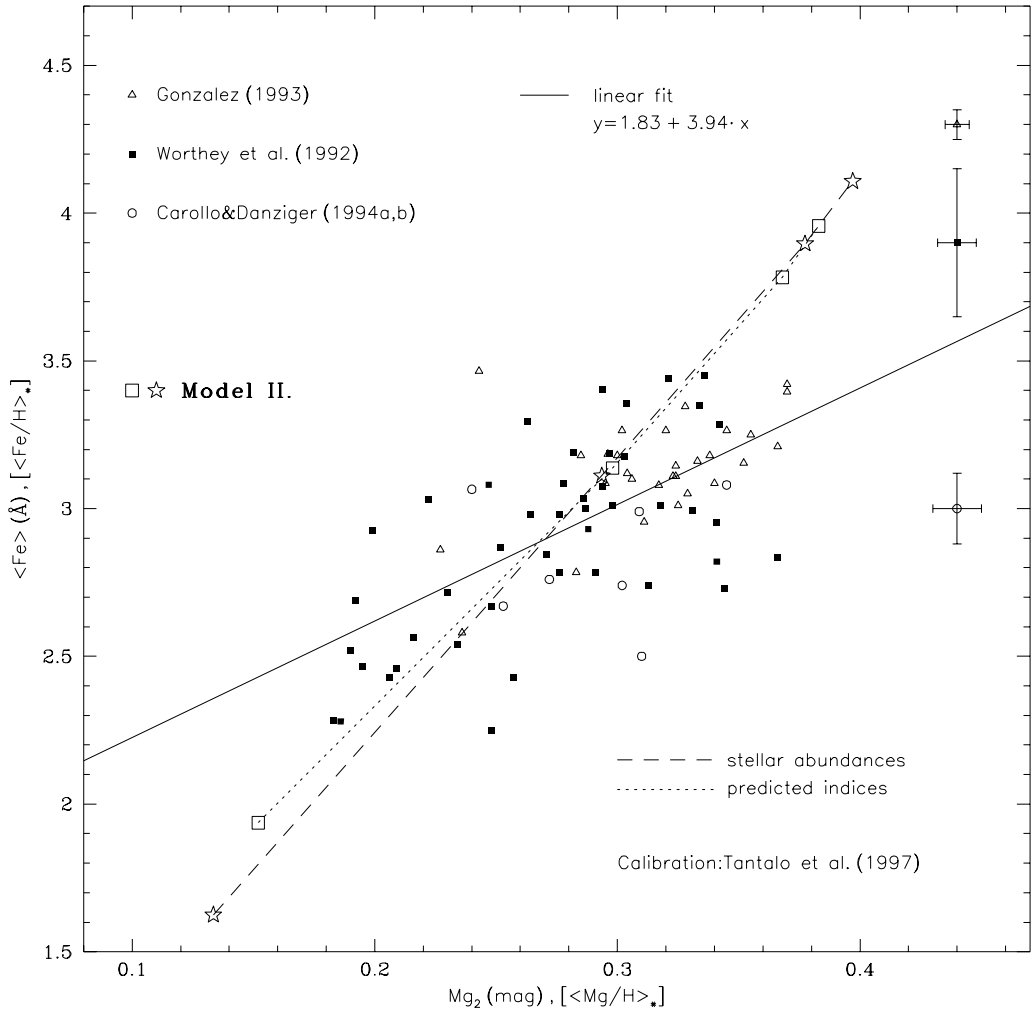
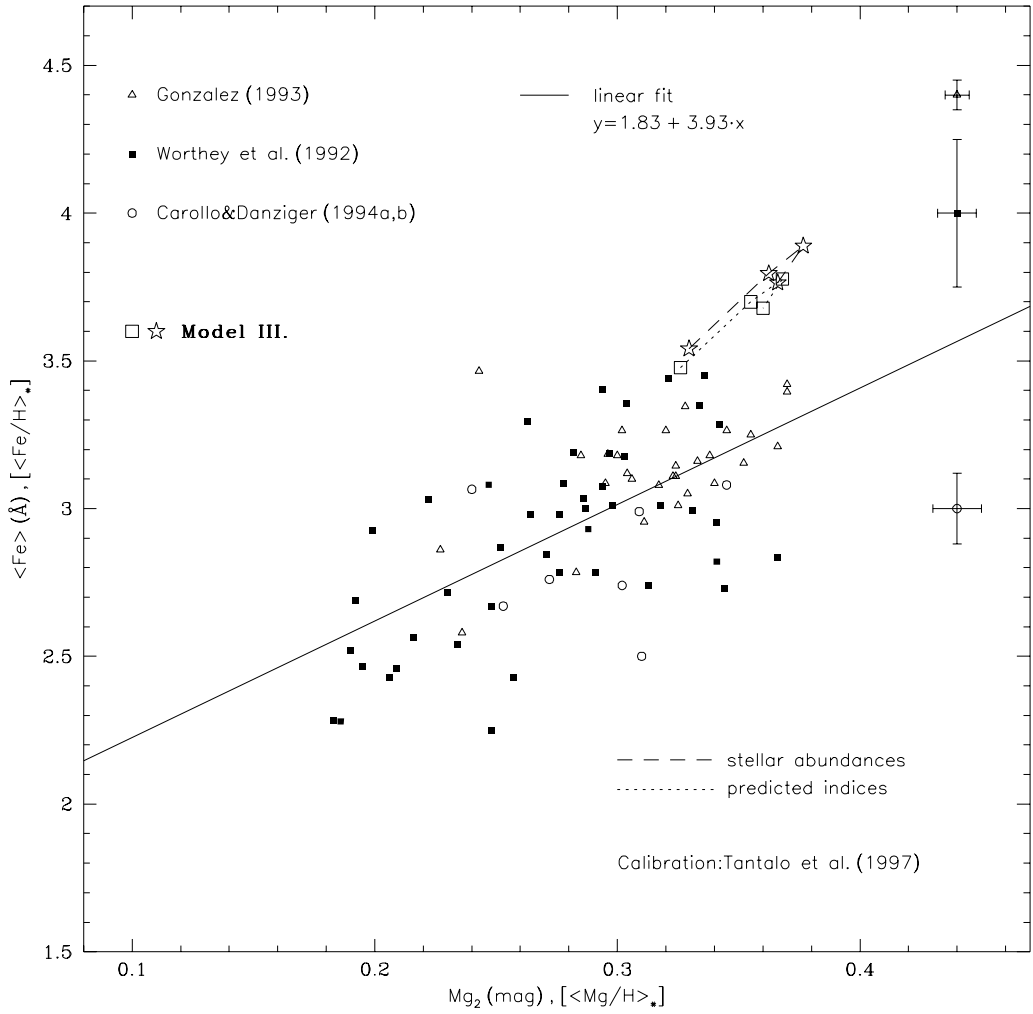
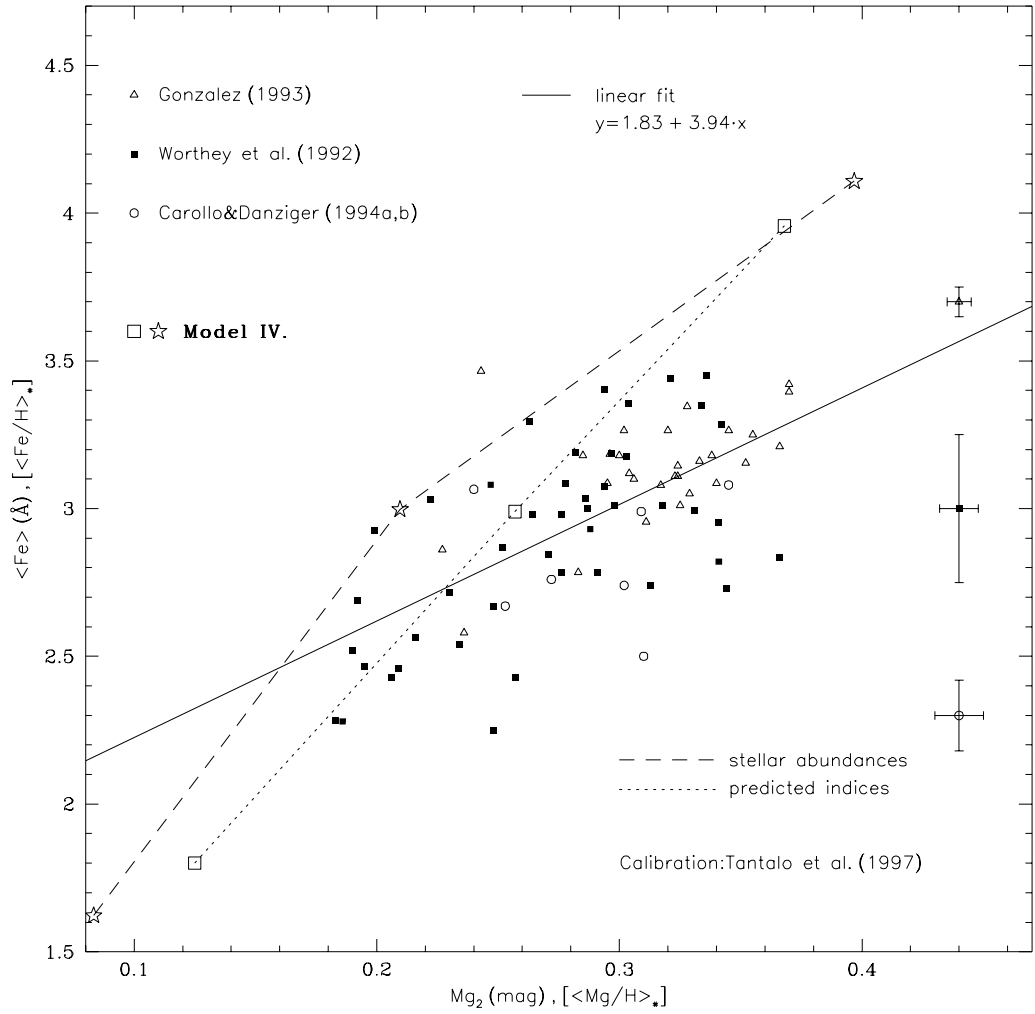
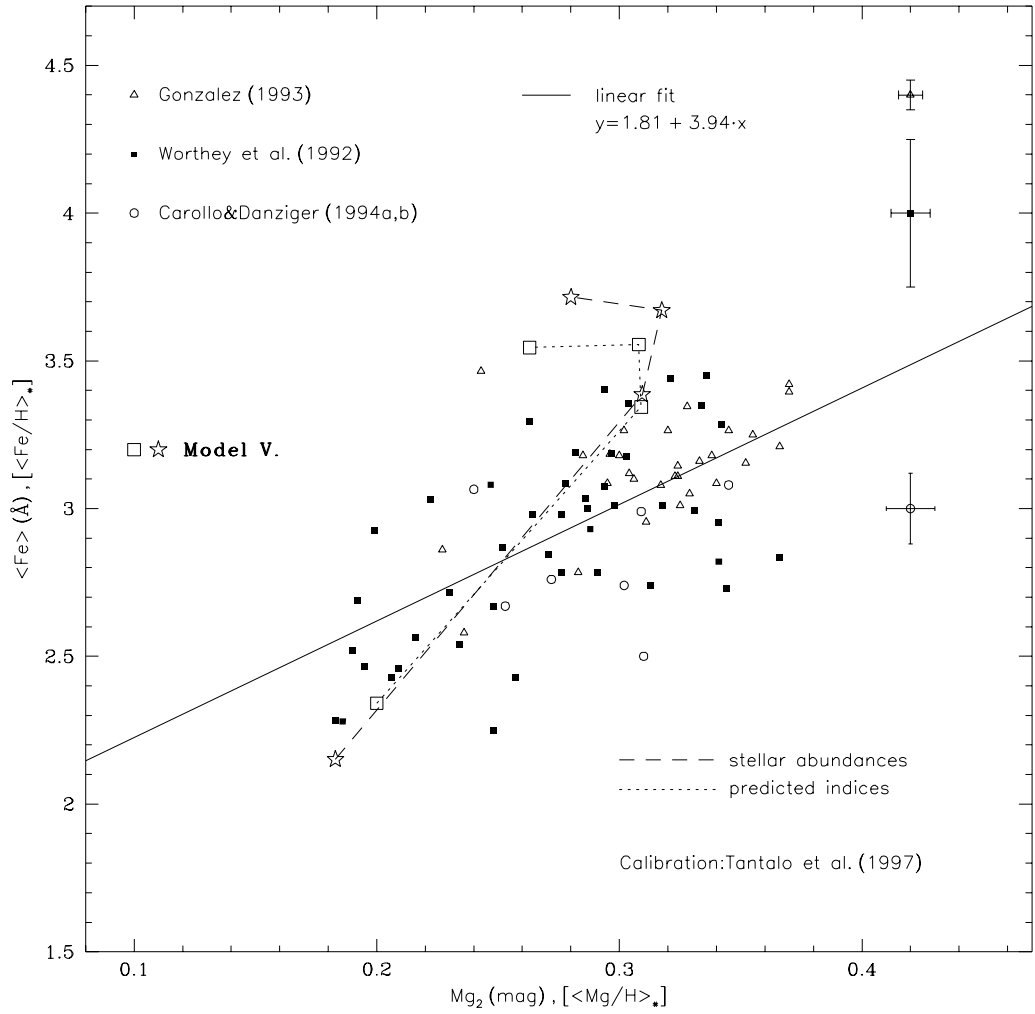
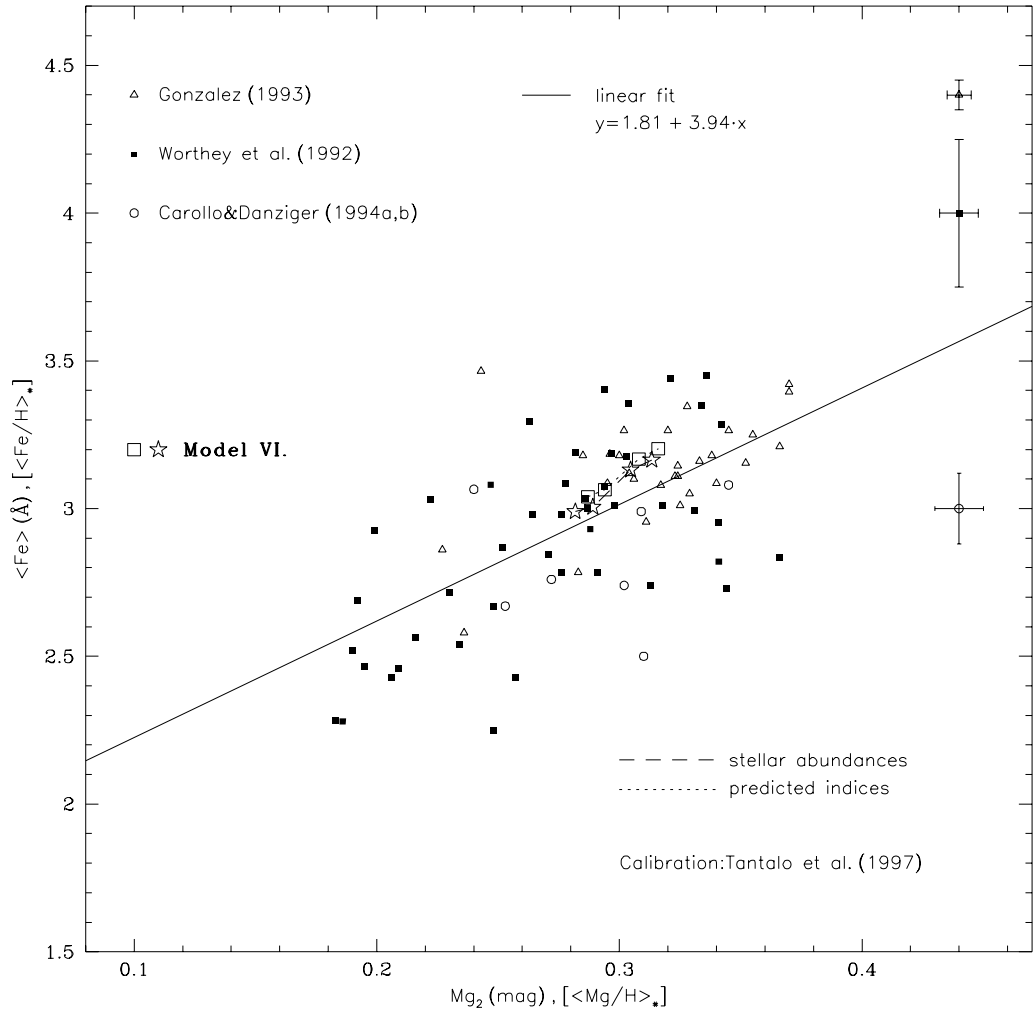
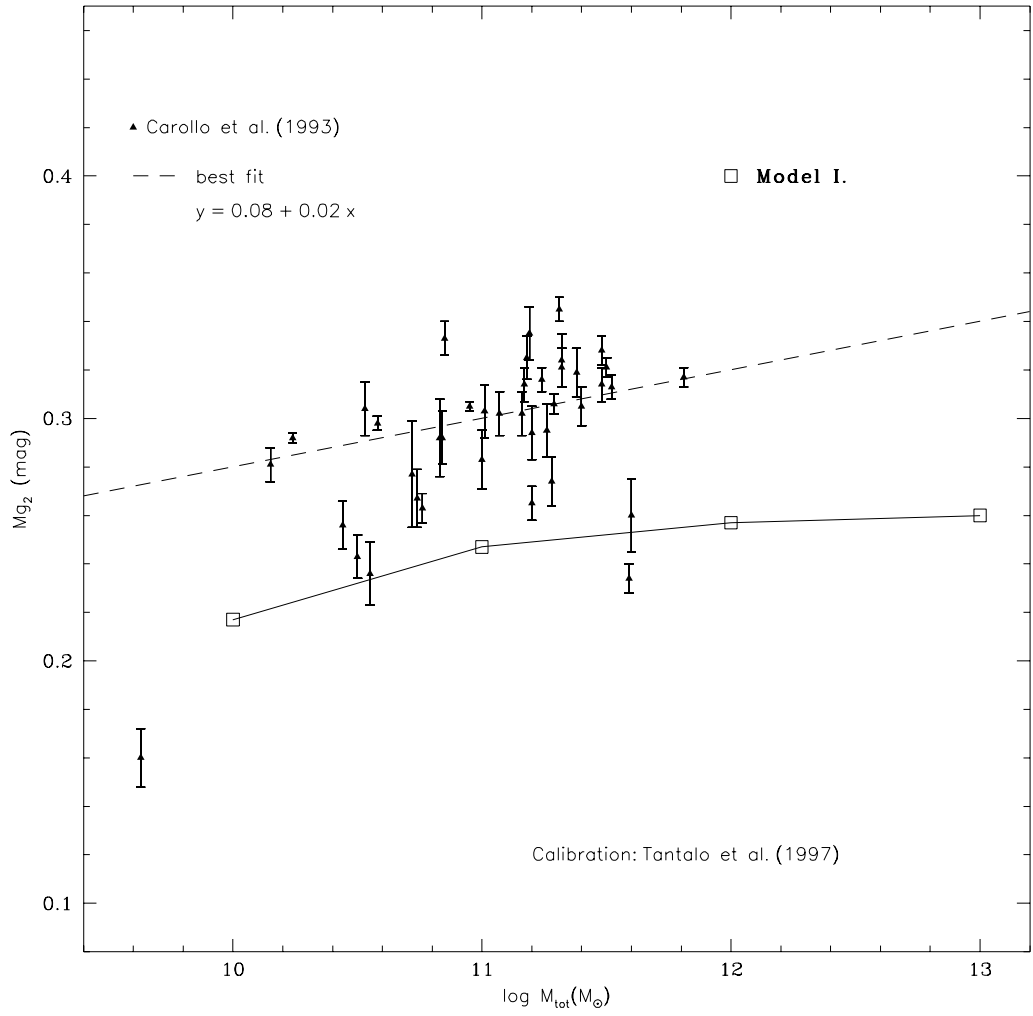
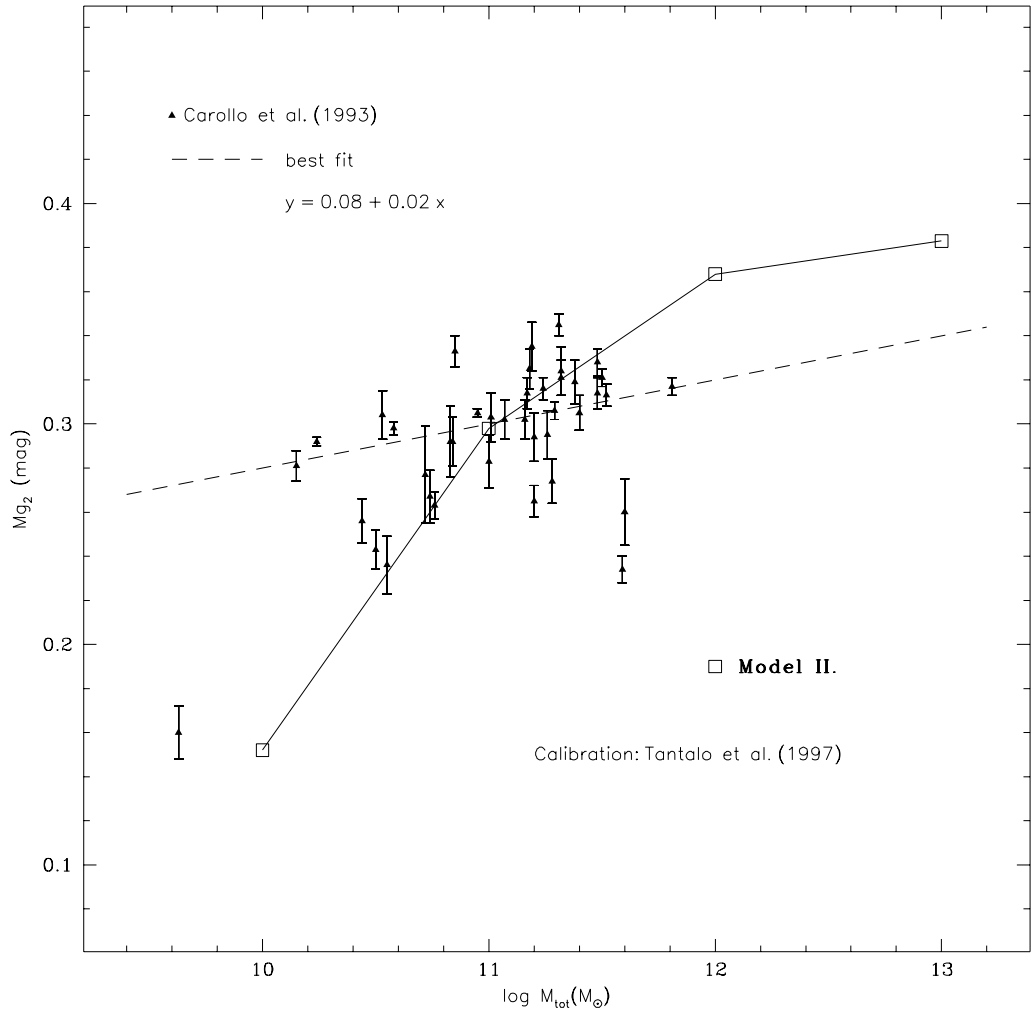
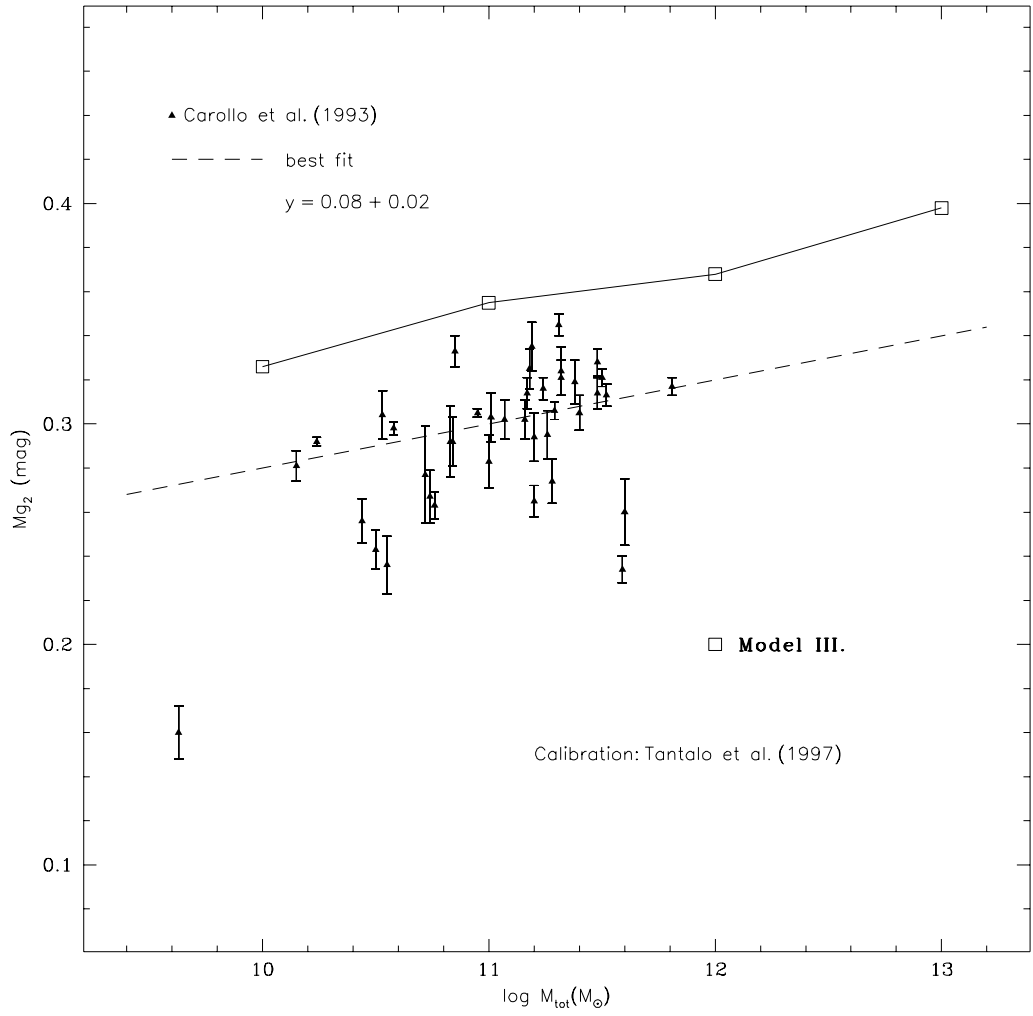
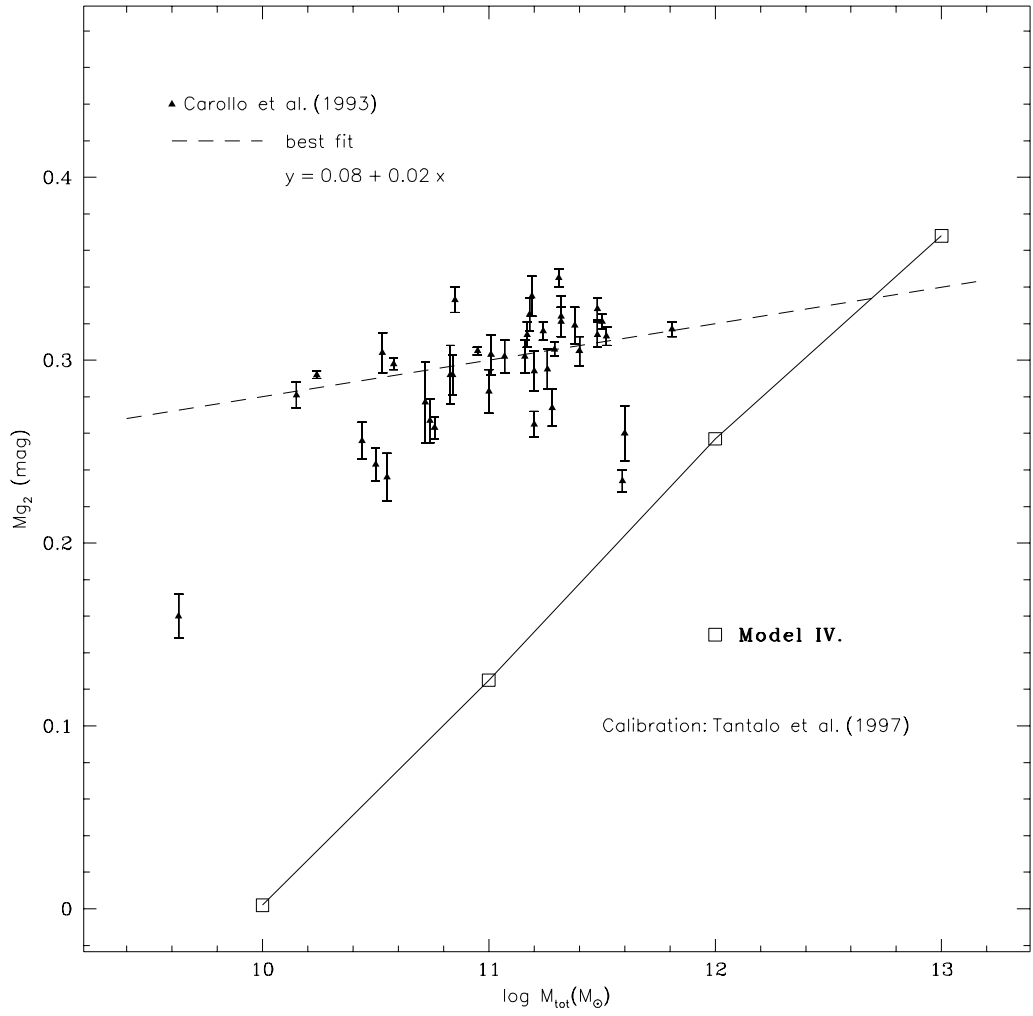
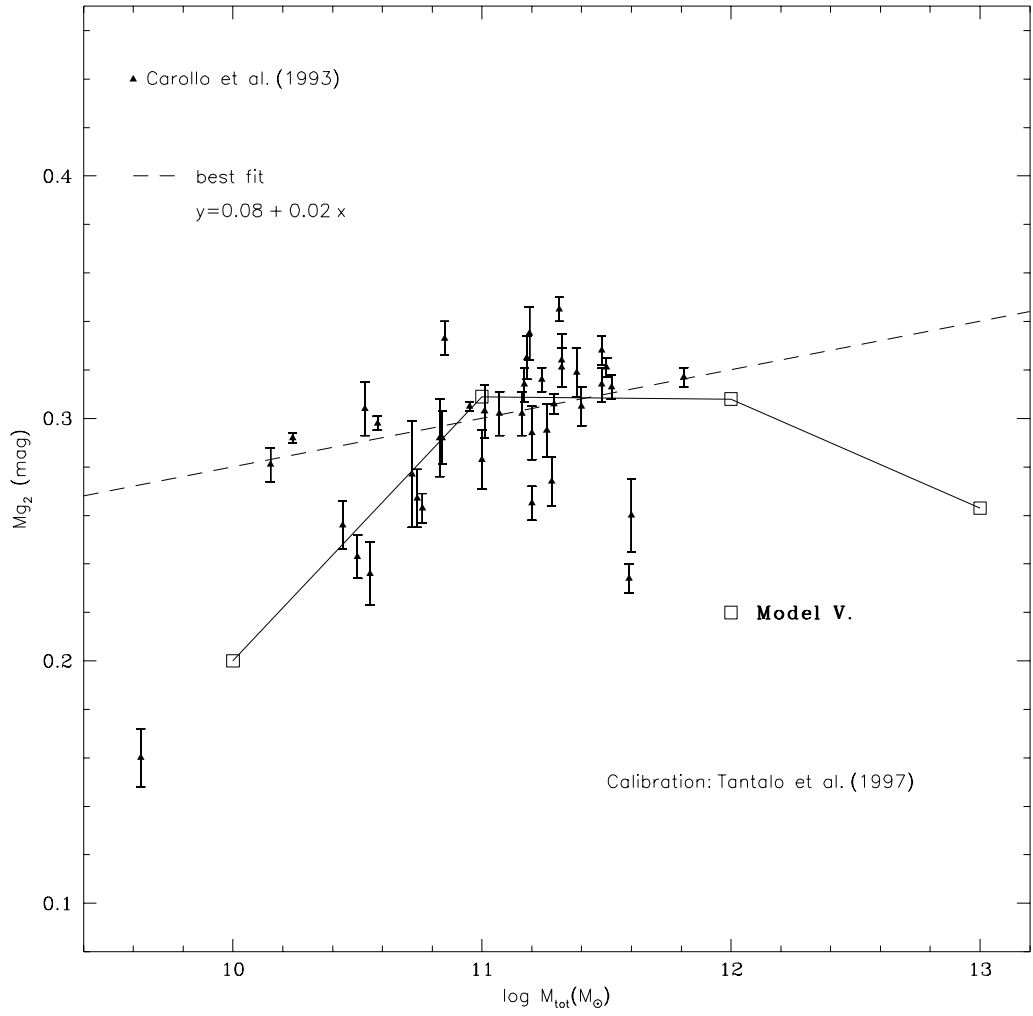
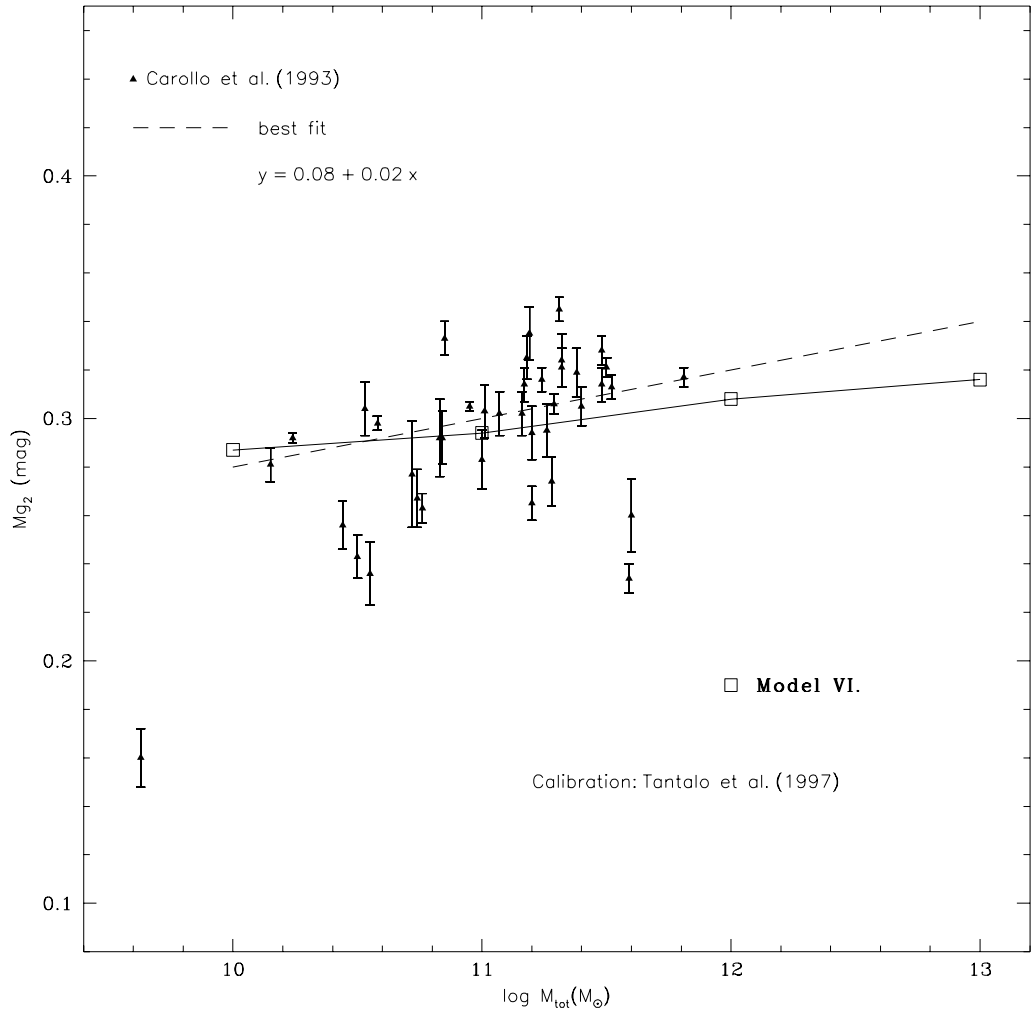
\end{document}